\pdfoutput=1
\documentclass[PAPER, atlasdraft=false, texlive=2025, UKenglish, cernpreprint, texmf, orcidlogo]{atlasdoc}
\usepackage{atlaspackage}
\usepackage{atlasbiblatex}

\usepackage{atlasphysics}
\graphicspath{{logos/}{figures/}}

\usepackage{ANA-HIGG-2023-12-PAPER-defs}
\usepackage{atlasunit}
\usepackage[inline,shortlabels]{enumitem}
\usepackage{multirow}

\AtlasTitle{Search for diphoton resonances in the 66 to 110 GeV mass range using $pp$ collisions at $\sqrt{s}=13$ TeV with the ATLAS detector}

\AtlasAbstract{%
A search is performed for light, spin-0 bosons decaying into two photons in the 66 to 110 GeV mass range, using \SI{140}{\per\fb} of proton-proton collisions at $\sqrt{s}=13$ TeV produced by the Large Hadron Collider and collected by the ATLAS detector.
Multivariate analysis techniques are used to define event categories that improve the sensitivity to new resonances beyond the Standard Model.
A model-independent search for a generic spin-0 particle and a model-dependent search for an additional low-mass Higgs boson are performed in the diphoton invariant mass spectrum.
No significant excess is observed in either search.
Mass-dependent upper limits at the \SI{95}{\%} confidence level are set in the model-independent scenario on the fiducial cross-section times branching ratio into two photons in the range of \SI{8}{\fb} to \SI{53}{\fb}.
Similarly, in the model-dependent scenario upper limits are set on the total cross-section times branching ratio into two photons as a function of the Higgs boson mass in the range of \SI{19}{\fb} to \SI{102}{\fb}.
}

\AtlasRefCode{HIGG-2023-12}

\PreprintIdNumber{CERN-EP-2024-166}

\arXivId{2407.07546}

\AtlasJournal{JHEP}
\AtlasJournalRef{JHEP 01 (2025) 053}
\AtlasDOI{10.1007/JHEP01(2025)053}

\hypersetup{pdftitle={ATLAS document},pdfauthor={The ATLAS Collaboration}}

\begin{document}

\maketitle

%

%
%

%
\newcommand{\AtlasCoordFootnote}{%
ATLAS uses a right-handed coordinate system with its origin at the nominal interaction point (IP)
in the centre of the detector and the \(z\)-axis along the beam pipe.
The \(x\)-axis points from the IP to the centre of the LHC ring,
and the \(y\)-axis points upwards.
Polar coordinates \((r,\phi)\) are used in the transverse plane,
\(\phi\) being the azimuthal angle around the \(z\)-axis.
The pseudorapidity is defined in terms of the polar angle \(\theta\) as \(\eta = -\ln \tan(\theta/2)\) and is equal to the rapidity
$ y = \frac{1}{2} \ln \left( \frac{E + p_z c}{E - p_z c} \right) $ in the relativistic limit.
Angular distance is measured in units of \(\Delta R \equiv \sqrt{(\Delta y)^{2} + (\Delta\phi)^{2}}\).}

\section{Introduction}
This paper presents a search for an additional light, spin-0 boson with an invariant mass ranging from \SIrange{66}{110}{\GeV} decaying into two photons, a signature that can arise in many theories of beyond the Standard Model (SM) physics.
These include two-Higgs-doublet models (2HDMs)~\cite{Celis:2013rcs,Fox:2017uwr} and next-to-2HDMs (N2HDMs)~\cite{Biekotter:2019kde}, next-to-minimal supersymmetric models (NMSSM)~\cite{Maniatis:2009re}, and models of supersymmetry that introduce pseudo-Nambu-Goldstone bosons (R-axions) through symmetry breaking~\cite{Bellazzini:2017neg}.
The introduction of an additional spin-0 boson can also be used to explain features observed in other experimental measurements.
For example, the excess of \GeV-scale gamma rays from the galactic centre can be explained if the additional spin-0 boson is a Higgs boson that acts as a scalar partner of dark matter~\cite{Cline:2019okt}.
Alternatively, if the additional spin-0 boson is an axion, even a weak coupling with the Higgs sector allows electroweak baryogenesis to explain the observed baryon asymmetry of the universe~\cite{Jeong:2018ucz}.

This analysis uses the full LHC proton-proton ($pp$) collision data sample at a centre-of-mass-energy of \SI{13}{\TeV} collected by the ATLAS detector (see Section~\ref{sec:detector}) during the years 2015--2018, corresponding to an integrated luminosity of \SI{140}{\per\fb} at a centre-of-mass energy of \SI{13}{\TeV}.
Both a model-independent search for a spin-0 particle ($X$) and a model-dependent search for an additional low-mass Higgs boson ($H$, assuming the production-mode cross-section times branching ratio into two photons as predicted by the SM at a given mass $m_{H}$) are performed.
In both cases, the assumption of a narrow-width resonance (NWA) is made and interference effects between the signal and background processes are neglected.
An additional search for larger-width signals is also performed in the model-independent search and considers ratios of decay widths $\Gamma_{X}$ to the mass $m_{X}$ of the spin-0 particle up to \SI{2.5}{\%}.
Additional signal widths are not considered for the model-dependent analysis due to the narrow decay width of SM-like Higgs bosons in the mass range considered here~\cite{LHCHiggsCrossSectionWorkingGroup:2013}.

Events that contain at least two photons (see Section~\ref{sec:sample}) are analysed for evidence of resonances in the diphoton invariant mass distribution.
The diphoton final state provides a clean experimental signature due to the excellent invariant mass resolution of the ATLAS detector.
In addition to the kinematic requirements, identification and isolation selections are applied to photons to reduce the impact of jet backgrounds and to ensure a high signal sensitivity (see Section~\ref{sec:object-selection}).
The event selection, described in Section~\ref{sec:event-selection}, uses photon transverse energy $E_{\textrm{T}}$\footnote{\AtlasCoordFootnote} selections that depend on the diphoton invariant mass $\mgg$ to suppress sculpting of the invariant mass distribution by the trigger selection.
A gradient boosted decision tree (BDT) is additionally used for photon--electron discrimination.

Multiple event categories (see Section~\ref{sec:classification}) are defined to maximise signal sensitivity.
The model-independent search for a spin-0 boson, $X$, splits the data into three categories based on whether the photons interact with nuclei in the detector that cause them to convert into a pair of electrons or not.
Alternatively,
the model-dependent search for an additional low-mass Higgs boson, $H$, employs a BDT to define three additional categories within each of the three photon-conversion categories, resulting in a total of nine categories.
This additional categorisation is only used for the model-dependent result due to the SM-like production-mode cross-sections assumed for the signal sample used to train the BDT.

The resonant \mgg signal distribution is modelled using analytic functions whose parameters are determined using Monte Carlo (MC) simulation, as described in Section~\ref{sec:signal}.
There are two main components of the background:
\begin{enumerate*}[1)]
\item the non-resonant \gamgam, \gamjet, and \jetjet processes that are henceforth referred to as the \emph{continuum background}, where ``j'' refers to a jet misidentified as a photon, and
\item the resonant \emph{Drell--Yan} (DY) dielectron processes (mainly $Z \rightarrow ee$ events) in which both the electrons are misidentified as photons.
\end{enumerate*}
The \mgg distributions of both the background components are described by analytic functions determined from MC simulation and data-driven background estimations, further described in Section~\ref{sec:background}.
The uncertainty in the continuum background due to limited data and MC simulated events is reduced using a Gaussian Process regression~\cite{GPML}.
The DY background affects prompt photons that convert to two electrons (converted photons) much more than those that do not (unconverted photons), hence a significant gain is obtained by splitting the analysis into separate conversion categories.

The final background \mgg shape parameters, background yield, and potential signal yield are obtained from a fit to the diphoton invariant mass distribution in data.
The \mgg region \SIrange{62}{120}{\GeV} is chosen to minimise the systematic uncertainty in the background model.
A search for hypothetical signal peaks in the range of \SIrange{66}{110}{\GeV} is performed, which ensures that there is enough data to constrain the background model both above and below the signal peak.
The resulting $p$-value scans are presented in Section~\ref{sec:results}.
Since no significant excess is observed, limits are set on the cross-section times branching ratio within a fiducial volume defined at particle-level in the model-independent search for $X$ and on the cross-section times branching ratio in the model-dependent search for $H$.

A previous search by the ATLAS Collaboration, using \SI{20.3}{\per\fb} of data at \SI{8}{\TeV}~\cite{HIGG-2014-04}, found no significant excesses.
Previous searches by the CMS Collaboration, using \SI{19.7}{\per\fb} of data at \SI{8}{\TeV} combined with \SI{35.9}{\per\fb} of data at \SI{13}{\TeV}~\cite{CMS-HIG-17-013}, and later with \SI{132.2}{\per\fb} of data at \SI{13}{\TeV}~\cite{CMS-HIG-20-002}, observed a maximal excess with a local (global) significance relative to the SM prediction of \SI{2.8}{$\sigma$} (\SI{1.3}{$\sigma$}) at \SI{95.3}{\GeV} and \SI{2.9}{$\sigma$} (\SI{1.3}{$\sigma$}) at \SI{95.4}{\GeV}, respectively.
%

%


%
\section{ATLAS detector}
\label{sec:detector}

The ATLAS experiment~\cite{PERF-2007-01} at the LHC is a multipurpose particle detector
with a forward--backward symmetric cylindrical geometry and a near \(4\pi\) coverage in solid angle.
It consists of an inner tracking detector surrounded by a thin superconducting solenoid
providing a \qty{2}{\tesla} axial magnetic field, electromagnetic and hadronic calorimeters, and a muon spectrometer.
The inner tracking detector covers the pseudorapidity range \(|\eta| < 2.5\).
It consists of silicon pixel, silicon microstrip, and transition radiation tracking detectors.
Lead/liquid-argon (LAr) sampling calorimeters provide electromagnetic (EM) energy measurements
with high granularity within the region \(|\eta|< 3.2\).
A steel/scintillator-tile hadronic calorimeter covers the central pseudorapidity range (\(|\eta| < 1.7\)).
The endcap and forward regions are instrumented with LAr calorimeters
for EM and hadronic energy measurements up to \(|\eta| = 4.9\).
The muon spectrometer surrounds the calorimeters and is based on
three large superconducting air-core toroidal magnets with eight coils each.
The field integral of the toroids ranges between \num{2.0} and \qty{6.0}{\tesla\metre}
across most of the detector.
The muon spectrometer includes a system of precision tracking chambers up to \(|\eta| = 2.7\) and fast detectors for triggering up to \(|\eta| = 2.4\).
The luminosity is measured mainly by the LUCID--2~\cite{LUCID2} detector, which is located close to the beampipe.
A two-level trigger system is used to select events~\cite{TRIG-2016-01}.
The first-level trigger is implemented in hardware and uses a subset of the detector information
to accept events at a rate below \qty{100}{\kHz}.
This is followed by a software-based trigger that
reduces the accepted event rate to \qty{1}{\kHz} on average
depending on the data-taking conditions.
A software suite~\cite{SOFT-2022-02} is used in data simulation, in the reconstruction
and analysis of real and simulated data, in detector operations, and in the trigger and data acquisition
systems of the experiment.


%
%
%
%
%
%
%
%
%
%

%
\section{Data and simulated event samples}
\label{sec:sample}

The ATLAS detector was used to collect \sqsthirt $pp$ collisions from the 2015--2018 LHC running periods, corresponding to an integrated luminosity of \SI{139.5\pm1.2}{\per\fb}~\cite{DAPR-2021-01} after data-quality requirements~\cite{DAPR-2018-01}.
The uncertainty in the integrated luminosity is obtained using the LUCID-2 detector~\cite{LUCID2} for the primary luminosity measurements, complemented by measurements using the inner detector and calorimeters.
The data were recorded using diphoton triggers that required two EM clusters with transverse energies \et
above a certain threshold and satisfying identification criteria based on variables describing the shape of the EM showers in the calorimeter (hereafter called \emph{shower shapes})~\cite{EGAM-2021-01}.
In the 2015 and the first portion of 2016 data taking, the \et threshold was \SI{20}{\GeV}, while in the remainder of 2016 data taking $\et > \SI{22}{\GeV}$ was required.
During 2017 and 2018 data taking, the \et threshold was reverted to \SI{20}{\GeV}, however an additional requirement on the sum of transverse energy around the photon candidate was applied~\cite{TRIG-2018-05}.

Simulated event samples are used to study signal and background processes, and to determine the analytic functions used to model both.
The search itself is performed by using the data to determine the parameters of the analytic functions (see Section \ref{sec:background}).
Interference effects between the resonant signal and all background processes are expected to be small for the signal widths considered here~\cite{ATL-PHYS-PUB-2016-009} and are neglected.

Background events containing two photons with associated jets were simulated with the \SHERPA 2.2.4~\cite{Gleisberg:2008ta,Bothmann:2019yzt} event generator.
Matrix elements were calculated with up to three partons at next-to-leading order (NLO) in quantum chromodynamics (QCD)~\cite{Siegert:2016bre}, and merged with the \SHERPA parton shower~\cite{Schumann:2007mg} according to the ME+PS@NLO prescription~\cite{Hoeche:2012yf}.
The CT10 parton distribution function (PDF) set~\cite{Lai:2010vv} was used in conjunction with dedicated parton shower tuning developed by the \SHERPA authors.
Events containing $Z$ bosons decaying into electron pairs were generated at NLO in QCD using \POWHEGBOX v2~\cite{Alioli:2010xd,Alioli:2008gx} interfaced to the \PYTHIA 8.186~\cite{Sjostrand:2007gs} parton shower model and the CT10 PDF set was used.
The AZNLO set of tuned parameters for the underlying event~\cite{STDM-2012-23} was used, with the \CTEQ[6L1] PDF set~\cite{Pumplin:2002vw}.
Additional events containing $Z$ bosons were also simulated with the \SHERPA 2.2.1 event generator, for comparison with \POWHEGBOX v2.

To better study electrons reconstructed as photons, single-electron and single-photon MC event samples were simulated with a pile-up profile corresponding to the Run 2 data sample.
These single-particle samples were generated with transverse energy distributions covering the range from \SI{5}{\GeV} to \SI{3}{\TeV}.

The signal samples assume a SM Higgs boson at different mass values, and were generated at NLO in QCD using \POWHEGBOX v2 interfaced to the \PYTHIA 8.186 parton shower model using the AZNLO set of tuned parameters, for the gluon--gluon fusion (ggF) and vector-boson fusion (VBF) production modes.
Samples were also simulated with the \PYTHIA 8.186 event generator using the A14 set of tuned parameters~\cite{ATL-PHYS-PUB-2014-021}, assuming the production of a Higgs boson in association with a \Wboson boson (WH), \Zboson boson (ZH) or top-quark pair (t\=tH).
Simulated samples were produced for fixed values of the mass of the assumed resonance, spanning the range \SIrange{60}{120}{\GeV}.
Generally, the model-independent search for $X$ uses the ggF sample as the nominal signal model with an uncertainty calculated by taking the envelope created by the other production modes, while the model-dependent search for a low-mass $H$ combines all production modes for a given mass point assuming SM-like cross-sections.
All assume a narrow-width resonance of \SI{4}{\MeV} that is negligible compared with the experimental resolution, which ranges from \SIrange{0.9}{2.2}{\GeV} (see Section~\ref{sec:signal}).

The effects of additional $pp$ interactions in the same or neighboring bunch crossings (pile-up) were modeled by overlaying soft QCD processes simulated with \PYTHIA 8.186 using the A2 set of tuned parameters~\cite{ATL-PHYS-PUB-2012-003} and the MSTW2008LO PDF set~\cite{Martin:2009iq}.
The simulated events are weighted to reproduce the distribution of average number of individual $pp$ interactions per bunch crossing and the distribution of the primary vertex $z$-position observed in data (pile-up reweighting).
All generated events were propagated through a detailed simulation of the ATLAS detector~\cite{SOFT-2010-01} based on \GEANT~\cite{Agostinelli:2002hh}.
The fast detector simulation used for background events containing two photons used a parameterisation of the performance of the calorimeters~\cite{ATL-PHYS-PUB-2010-013}.
%


%
\section{Event reconstruction}
\label{sec:object-selection}

The event reconstruction is similar to the one described in Ref.~\cite{HIGG-2019-13}.
Photon candidates are reconstructed from topological clusters of energy deposited in the EM calorimeter (calorimeter clusters), and from charged-particle tracks and conversion vertices reconstructed in the ID.
Photon clusters without any corresponding track in the ID are considered unconverted.
Two opposite-charge tracks that form a vertex consistent with a massless particle and that both match to calorimeter clusters are considered converted photons; single-track vertices---essentially those without hits in the innermost ID layers---that match to a calorimeter cluster are also considered as converted photons~\cite{EGAM-2018-01}.
Photon conversion fractions vary from \SI{20}{\%} to \SI{65}{\%}, depending on $\eta$~\cite{EGAM-2018-01}.
Photons and electrons are required to fulfil identification criteria based on the shower shapes in the EM calorimeter~\cite{EGAM-2018-01}.
Residual differences between the average values of the shower-shape variables measured in data and simulation using $Z \rightarrow \ell\ell\gamma$ events are corrected by shifting the shower-shape distributions in the simulation~\cite{PERF-2017-02}.
The photon identification efficiency is determined as a function of \ET and $|\eta|$ and the efficiency increases with \ET from 70\% at \SI{20}{\GeV} to 90\% at \SI{50}{\GeV}~\cite{EGAM-2021-01}.

Photons are required to be in the high-precision EM calorimeter within the pseudorapidity interval $|\eta|<2.37$, excluding the transition region $1.37<|\eta|<1.52$ between the barrel and end-cap calorimeters.
Corrections to the energy of photon clusters is based on a multivariate regression algorithm optimised on simulated samples and scale factors derived extracted from data samples~\cite{EGAM-2021-02}.
The two candidates with the highest transverse energies, both satisfying $\ET>\SI{22}{\GeV}$, are retained.
This transverse energy requirement is slightly higher than the trigger threshold, for most of the data-taking periods, to mitigate the trigger efficiency turn-on effect.
A subsequent requirement on the ratio of photon \ET to the diphoton mass (see Section~\ref{sec:event-selection}) further raises the energy of selected photons.
Primary vertices are reconstructed using at least two good-quality tracks with $\pT > \SI{500}{\MeV}$~\cite{ATL-PHYS-PUB-2015-026}.
The photon candidates are used to select the diphoton vertex using a neural-network algorithm based on charged-particle tracks and primary vertex information, as well as the direction of the two photons measured in the calorimeters and ID~\cite{HIGG-2013-08}.
Once the diphoton vertex is selected, the direction of the two photon candidates is re-computed relative to this updated primary vertex.
This recomputation improves the \ET measurement of each photon candidate through its dependence on the photon candidate’s direction in $\eta$.
The updated energy measurement improves the diphoton invariant mass resolution by about 8\% for inclusive Higgs boson production relative to the default primary vertex selection~\cite{ATL-PHYS-PUB-2015-026}.

Electron candidates---used for studying the DY background process---are reconstructed by matching tracks in the inner detector with clusters of energy deposits in the EM calorimeter formed with the same algorithm as in the photon reconstruction~\cite{EGAM-2018-01}.
The tracks are required to be consistent with the diphoton vertex using their longitudinal ($z_0$) and transverse ($d_0$) impact parameters.
Electrons must also satisfy the same \ET and $\eta$ selection criteria as photons.

To improve the rejection of jets misidentified as photons, the candidates are required to be isolated using both the calorimeter and tracking detector information.
The calorimeter-based isolation variable \etiso is defined as the scalar sum of the \ET over positive-energy topological clusters~\cite{PERF-2014-07} within a radius $\Delta R = 0.2$ around the photon candidate, excluding the photon energy and correcting for pile-up and underlying-event contributions~\cite{STDM-2010-08,Cacciari:2008gn,Cacciari:2009dp}.
For both the candidates this variable is required to be below $0.065 \times \ET$, where \ET is the transverse energy of the photon.
The track-based isolation variable \ptiso is based on charged-particle tracks, and is defined as the scalar sum of the transverse momenta \pT over tracks within a radius $\Delta R = 0.2$ around the photon candidate.
Only tracks with $\pT>\SI{1}{\GeV}$ that are consistent with originating from the diphoton production vertex and that are not associated with a photon conversion vertex are used.
For both the candidates this variable is required to be below $0.05 \times \ET$.
Small differences between the average value of \etiso between data and simulation are corrected in the simulation.
The photon isolation efficiency---i.e., the fraction of photons fulfilling the identification requirement that also satisfy the isolation requirement---is determined using simulated samples and increases with \mgg from \SI{80}{\%} at \SI{62}{\GeV} to \SI{90}{\%} at \SI{120}{\GeV}.


%
\section{Event selection}
\label{sec:event-selection}

Only events containing at least one primary vertex candidate are considered.
The two selected photon candidates are used to define the diphoton invariant mass, \mgg, and only events in the mass range \SIrange{62}{120}{\GeV} are included in the analysis.
To avoid distortions in the diphoton invariant mass spectrum due to the kinematic turn-on effects from the trigger selection, each photon is required to satisfy $\ET/\mgg > 22/58 \approx 0.38$.
This particular value is chosen to maximise the signal efficiency while allowing the mass range \SIrange{62}{120}{\GeV} to be described more easily using monotonically decreasing analytic functions (see Section~\ref{sec:background}).

Following these selections, two significant background components are identified: \gamgam, \gamjet and \jetjet continuum backgrounds coming from QCD production, and $ee$ pairs coming from DY production.
Here, the \gamjet component includes events where either the leading or sub-leading object is a jet misidentified as a photon, and the \jetjet component includes events where both the objects are jets misidentified as photons.
The relative contribution of the continuum backgrounds varies by \mgg and conversion category, with roughly \SI{75}{\%} due to \gamgam, \SI{20}{\%} due to \gamjet, and \SI{5}{\%} due to \jetjet processes (see Section~\ref{sec:background}).
The contributions of events containing SM Higgs bosons or $W$ and $Z$ bosons produced in association with a photon are estimated by using MC simulated event samples and are found to be negligible.

\begin{figure}
\centering
\subfloat[]{
\includegraphics[width=0.5\textwidth]{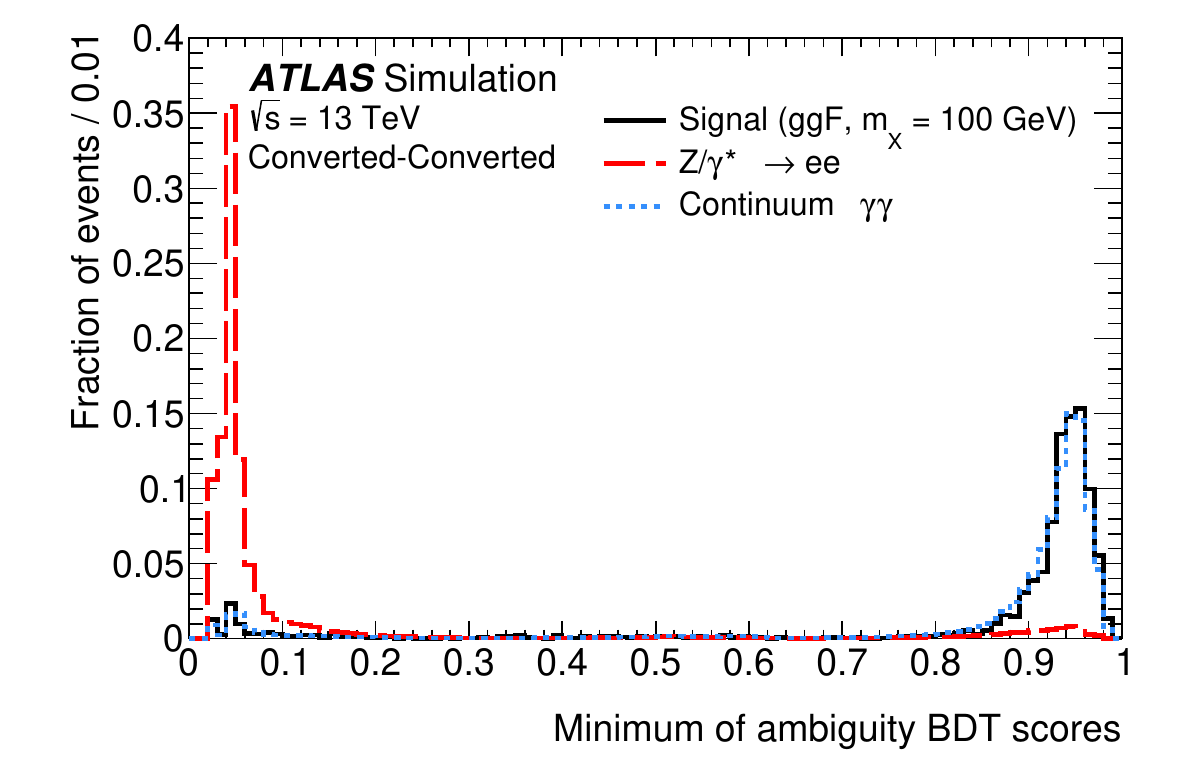}
}
\subfloat[]{
\includegraphics[width=0.5\textwidth]{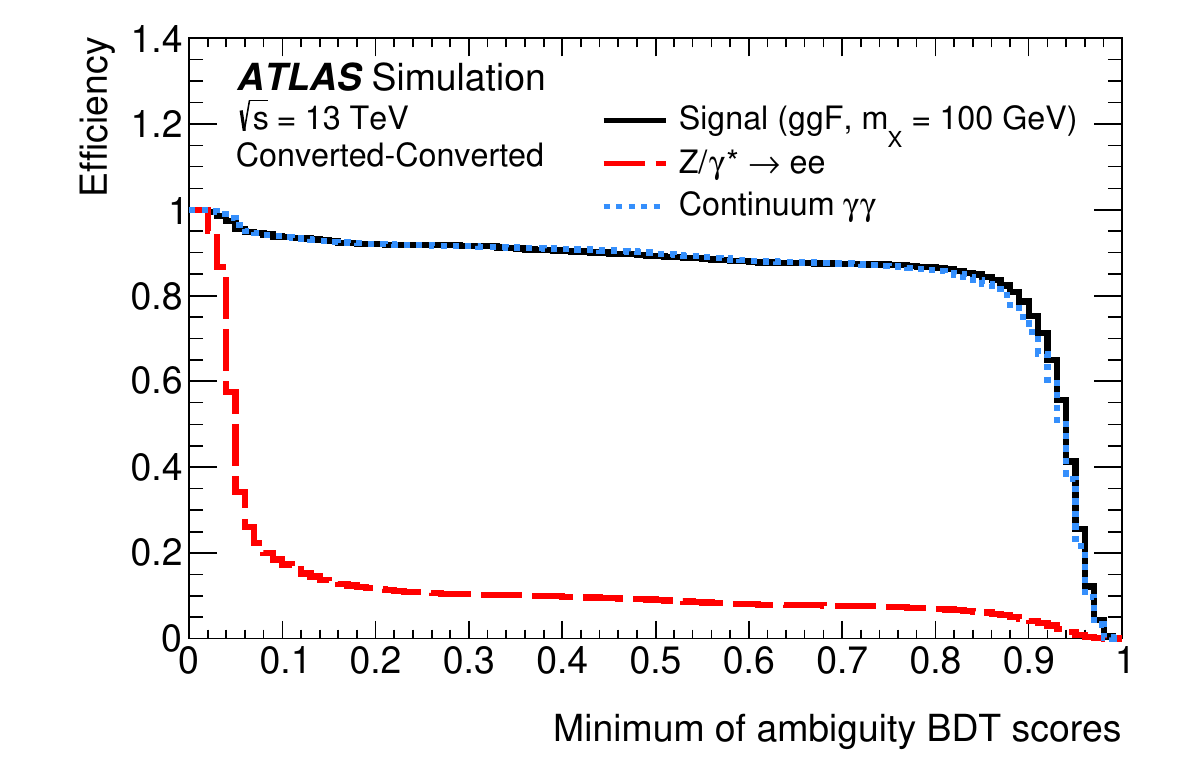}
}
\caption{
(a) Distribution of electron--photon ambiguity BDT scores constructed by taking the minimum score of the two photon candidates in simulated ggF $\mX = \SI{100}{\GeV}$ signal events (solid line), $Z/\gamma^{*} \rightarrow ee$ events (dashed line), and the continuum $\gamma\gamma$ background events (dotted line).
The distributions are scaled independently for illustrative purposes.
(b) Efficiency versus minimum requirement on the ambiguity BDT score, shown for the same samples.
Only events in which both photon candidates are converted are shown.
}
\label{fig:object-minBDT-score}
\end{figure}

To further reduce the number of background $ee$ events, in particular the events where a topological cluster is ambiguously reconstructed as both an electron and a photon~\cite{EGAM-2018-01}, a boosted decision tree (BDT) is developed using LightGBM~\cite{NIPS2017_6907}.
Photon candidates from single-photon MC samples are considered as signal in the training, while photon candidates from the single-electron MC samples are considered as background when training the BDT.
The model is trained using kinematic information related to the converted photon and detailed tracking information related to the electron and conversion of the photon candidate.
The full list of variables considered for this classifier are chosen to avoid introducing shapes in the background distribution that are difficult to model with analytic functions (see Section~\ref{sec:background}).
The output of this classifier (henceforth referred to as the \emph{ambiguity BDT}) is a score on the interval [0,1], where scores closer to 0 indicate an electron-like object and a score nearer to 1 indicate a photon-like object.
Unconverted and non-ambiguous photons are assigned a score of 1 and are not further classified.
The score of the ambiguity BDT is then evaluated for each of the two photon candidates.
In Figure~\ref{fig:object-minBDT-score}, the minimum score between the two candidates and the selection efficiency for the signal and background processes are shown, both as a function of ambiguity BDT score in events with two converted photon candidates and thus most likely to suffer from $ee$ backgrounds.
Requiring both the ambiguity BDT scores to be above 0.2 results in a signal selection efficiency above \SI{93}{\%} and a reduction of $ee$ backgrounds between \SI{65}{\%} to \SI{90}{\%}, where the largest reduction is obtained for events with two converted photons.

%
\section{Event categorisation}
\label{sec:classification}

As the $ee$ backgrounds predominantly impact events where both photons convert to a pair of electrons (see Section~\ref{sec:object-selection}), including distinct conversion categories results in a significant increase in sensitivity to new physics.
After the event selection from Section~\ref{sec:event-selection} is applied, events are categorized into those where:
\begin{enumerate*}[1)]
\item both photons remain unconverted (UU),
\item either one of the two photons convert (UC), or
\item both photons convert (CC).
\end{enumerate*}

\begin{figure}
\centering
\includegraphics[width=0.6\textwidth]{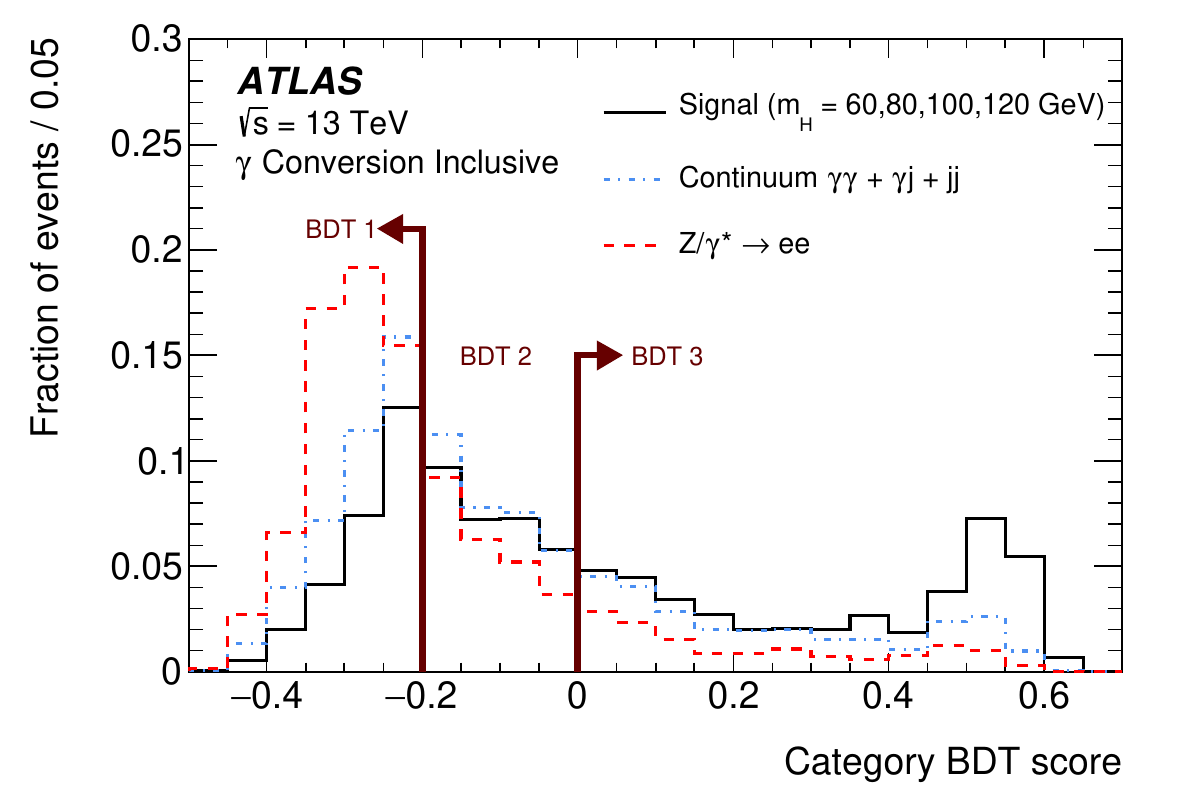}
\caption{
Distributions of the category BDT scores for the merged SM-like Higgs boson considering all production modes (ggF, VBF, t\=tH, WH, ZH), the diphoton ($\gamma\gamma$) and reducible backgrounds ($\gamma$j, jj) continuum,
and the simulated $Z\rightarrow ee$ background prediction.
The reducible background components are derived from dedicated data control regions where the photon identification and isolation requirements are inverted.
The merged signal contains signals generated for $m_{H} = $ 60, 80, 100, and \SI{120}{\GeV} and each MC sample is weighted according to the SM-like Higgs boson cross-section.
Photons from all three conversion categories are used in the BDT training.
The merged signal and backgrounds are separately normalised to unity.
The vertical lines and arrows at category BDT scores of $-0.2$ and 0 define the categorisation used in this analysis.
Events with category BDT scores below $-0.2$ are in BDT 1, events with category BDT scores between $-0.2$ and 0 are in BDT 2, and events with category BDT scores above 0 are in BDT 3.
}
\label{fig:category-bdt-score}
\end{figure}

For the model-dependent result, in addition to the conversion categories, events are separated between the continuum background and low-mass Higgs boson processes by another BDT, henceforth referred to as the \emph{category BDT}.
The training of this BDT is performed using the adaptive boosting (AdaBoost) algorithm~\cite{Freund:1997xna}, with a boosting parameter of 0.5, designed with the TMVA toolkit~\cite{TMVA}.
SM-like assumptions on the production-mode cross-sections are used for the model-dependent result.
Eight input variables are considered and are given here ranked in order of variable importance in the training: the cosine of the difference in azimuthal angle between the two photons, the ratio $\ET/\mgg$ of each photon, the $\eta$ of the each photon, the minimum of the two ambiguity BDT scores, and the ambiguity BDT scores of each of the two photons.
Photon candidates from simulated samples with $m_{H} = $ 60, 80, 100, and \SI{120}{\GeV}---where the different masses and production modes are weighted assuming the SM-like cross-sections for a Higgs boson at the specified $m_{H}$---are considered as signal (S) and simulated diphoton events are considered as background (B) when training this BDT.
For simplicity, the photon candidates from all three conversion categories are used in the BDT training.
The resulting category BDT score is shown in Figure~\ref{fig:category-bdt-score}, for signal and background processes.
The category BDT score is used to define three sub-categories within each of the three conversion categories, labelled as \SIrange{1}{3}{} in increasing order of expected signal compared to background.
The nine categories are defined based on the combined conversion and BDT category as UU1, UU2, UU3, UC1, UC2, UC3, CC1, CC2, and CC3.
The number of categories and the corresponding boundaries are optimised based on the expected signal-to-background significance, quantified by $S/\sqrt{B}$, while requiring that there are a sufficient number of simulated events for determining the background model.
Each sub-category is required to contain at least 20\% of the total number of diphoton background events in a given photon conversion category.
While small differences in the optimal BDT score boundaries are found for different categories, for simplicity, the same boundaries in BDT score of $-0.2$ and 0 are used for all events.
A list of each category and its requirements can be found in Table~\ref{table:categories_selections}.
\begin{table}
\caption{The selection requirements and names of each category in the model-independent and model-dependent analyses.}
\begin{center}
\begin{tabular}{cl} \toprule
Category & Selection Requirement \\ \midrule
\multicolumn{2}{l}{\underline{Model-Independent categories}} \\
UU & 2 unconverted photons \\
UC & 1 converted photon and 1 unconverted photon \\
CC & 2 converted photons \\
\midrule
\multicolumn{2}{l}{\underline{Model-Dependent categories}} \\
UU1 & UU and category BDT score < $-0.2$ \\
UU3 & UU and category BDT score [$-0.2$,0) \\
UU3 & UU and category BDT score >= 0 \\
UC1 & UC and category BDT score < $-0.2$ \\
UC2 & UC and category BDT score [0.2,0) \\
UC3 & UC and category BDT score >= 0 \\
CC1 & CC and category BDT score < $-0.2$ \\
CC3 & CC and category BDT score [$-0.2$,0) \\
CC3 & CC and category BDT score >= 0 \\
\bottomrule
\end{tabular}
\end{center}
\label{table:categories_selections}
\end{table}

The category BDT classification is found not to cause significant distortions of the diphoton invariant mass distribution for any category.
The expected signal and background yields for a signal with $\mH = \SI{90}{\GeV}$ are given in Table~\ref{tab:diphoton_bdt_yield}.

\begin{table}
\centering
\caption{
The expected number of signal events, fractions of each Higgs boson production mode, and the number of background events per \GeV\ at $\mgg = \SI{90}{\GeV}$ for each BDT category, and for all three photon conversion categories together.
The per \GeV\ binning corresponds to approximately $1\sigma$ of the mass resolution at $\mgg = \SI{90}{\GeV}$.
The background events are extracted from the background-only fit to the data and the ``Total'' category includes the number of Drell-Yan events (DY) that are also shown separately.
The BDT categories are defined as follows: events with category BDT scores below $-0.2$ are in BDT category 1, events with category BDT scores between $-0.2$ and 0 are in BDT category 2, and events with category BDT scores above 0 are in BDT category 3.
}
\begin{tabular}{c|cccccccccc}
\hline\hline
&  & \multicolumn{6}{c}{SM-like Higgs boson ($m_{H} = \SI{90}{\GeV}$)} &  & \multicolumn{2}{c}{Background} \\
\cline{3-8} \cline{10-11}
BDT Category &  & Total & ggF & VBF & WH & ZH & t\=tH  &  & Total & DY \\
&  &  & [\%] & [\%] & [\%] & [\%] & [\%] &  & [$\GeV^{-1}$] & [$\GeV^{-1}$] \\
\hline
1     &  & 741  & 97.1  & \phantom{0}1.2   & 1.0 & 0.6 & 0.1 & & 18877  & 2179 \\
2     &  & 942  & 93.4  & \phantom{0}2.9   & 2.1 & 1.2 & 0.4 & & 14014 & 713  \\
3     &  & 1187 & 72.4  & 13.5  & 6.7 & 4.0 & 3.4 & & \phantom{0}6522  & 294  \\
\hline
Total &  & 2870 & 85.7  & \phantom{0}6.8   & 3.7 & 2.2 & 1.6 &  & 39413 & 3186 \\
\hline\hline
\end{tabular}
\label{tab:diphoton_bdt_yield}
\end{table}


%
\section{Signal model}
\label{sec:signal}

\begin{figure}[bp]
\centering
\subfloat[]{\includegraphics[width=0.49\textwidth]{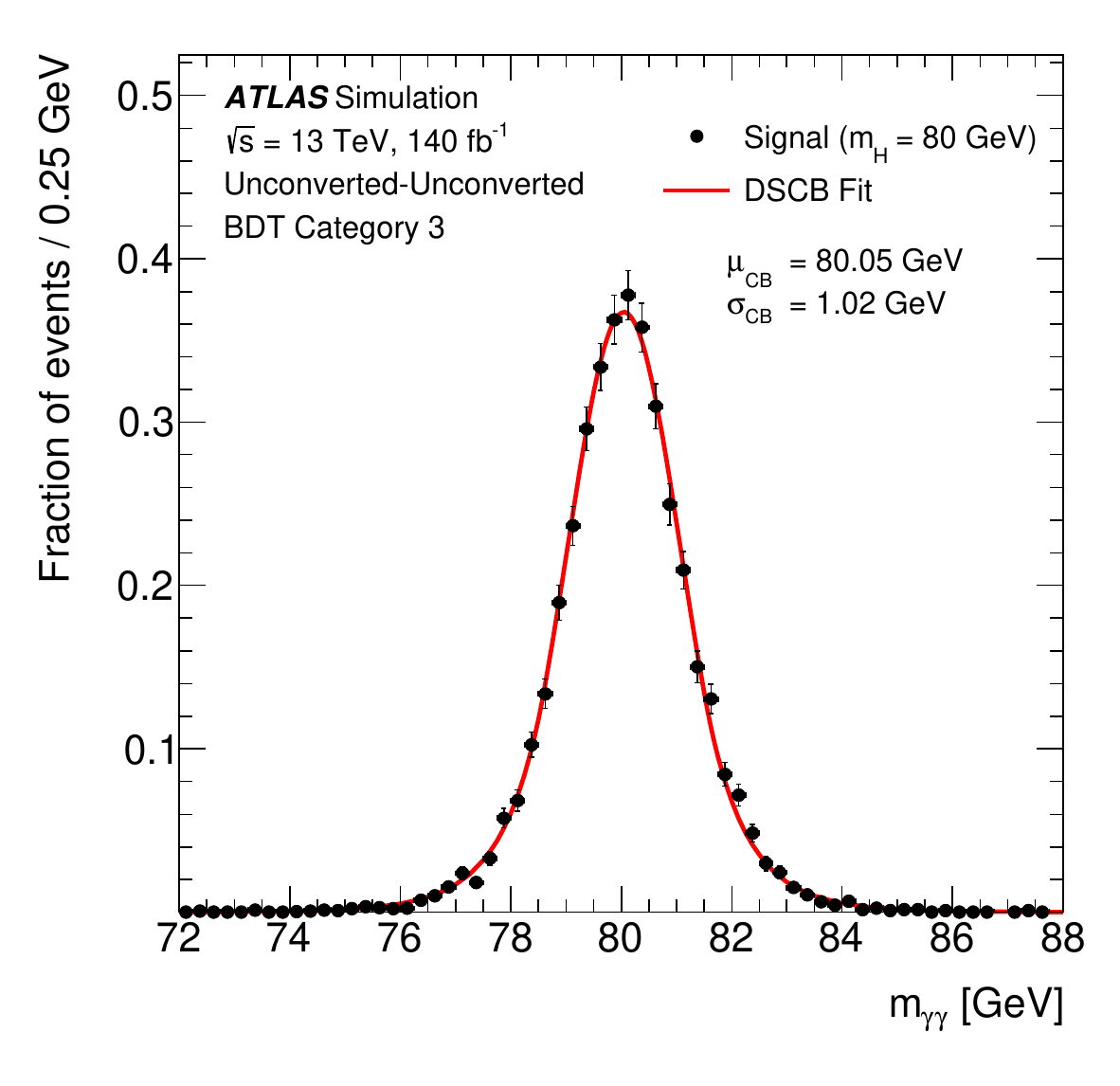}}
\subfloat[]{\includegraphics[width=0.49\columnwidth]{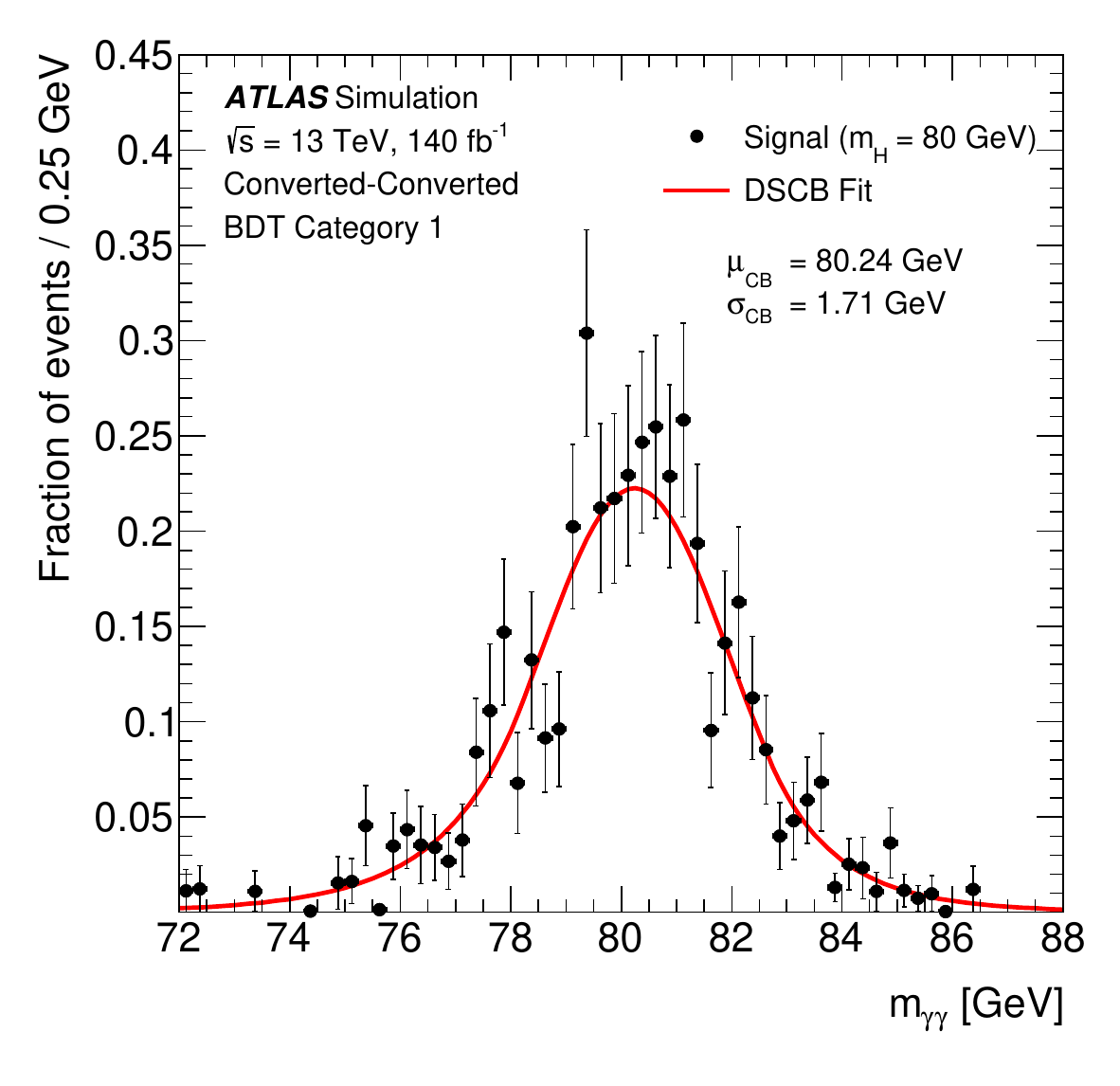}}
\caption{Simulated diphoton invariant mass distribution of a narrow-width signal
particle $H$ of mass \SI{80}{\GeV} (points) in the (a) UU3 and (b) CC1 categories, overlaid with the DSCB
function resulting from the signal model parameterisation (line).
The error bars on the simulated data points indicate the statistical uncertainties.
An arbitrary normalisation is used for illustration purposes.
}
\label{fig:signal}
\end{figure}

The \mgg distribution of the signal, assumed to have a narrow decay width relative to the mass resolution, is modelled using a double-sided Crystal Ball (DSCB) function, composed of a Gaussian core with power-law tails~\cite{Oreglia:1980cs,HIGG-2016-08}.
Each parameter is determined in a fit to the fixed-mass simulated samples, and is parameterised as a linear function of the resonance mass separately for each conversion category.
For the model-independent search for a spin-0 scalar $X$ only the ggF production mode is considered; for the model-dependent search for a low-mass $H$ all production modes for a given mass value are combined assuming SM-like cross-sections~\cite{deFlorian:2016spz}.
The width of the Gaussian core, which is entirely determined by detector resolution, ranges from \SIrange{0.9}{2.2}{\GeV} depending on the resonance mass and analysis category.
Good agreement between the signal model fit and the simulated \mgg distribution is found, with reduced chi-square values generally between 0.9 and 1.4.
Examples fits are illustrated in Figure~\ref{fig:signal} for the UU3 and CC1 categories.

In addition to the narrow width signals considered above, a search for signals with larger widths is also performed in the model-independent analysis.
The large width signal distributions for specific values of the mass and width are derived by convolving the detector resolution with the predicted line shape defined in the MC simulation of the ggF process.
The detector resolution is modelled using the DSCB function described above and the signal line shapes are comprised of a Breit-Wigner (BW) function of width $\Gamma_{\textrm{X}}$ combined with a mass-dependent gluon–gluon luminosity functional form.
The large width signal hypotheses considered in this analysis cover the range $\SI{0}{\%} \leq \Gamma_\textrm{X}/\mX \leq \SI{2.5}{\%}$.
To ensure that sufficient side bands are available for the largest width signals, only signal masses in the range \SIrange{75}{105}{\GeV} are considered for the large-width analysis.


%
\section{Background estimate}
\label{sec:background}

The two main components of the background, the continuum and the resonant DY, are estimated separately in each category.
In both cases, a data-driven approach is used to describe the normalisation and shape of their \mgg distributions.
The continuum background is fitted on data, with the normalisation and function parameters free.
For the DY background, both the shape and normalisation are fitted, but the corresponding parameters are constrained using information from $Z\rightarrow ee$ decays, as described below.

The continuum background is predominately composed of events from $\gamgam$ but also $\gamjet$ and $\jetjet$ production processes.
To evaluate its composition,
the two-dimensional side bands method described in Ref.~\cite{STDM-2012-05} is used to simultaneously extract the fraction of \gamgam, \gamjet, and \jetjet events (including fake photons coming from DY electrons).
The inputs to this method are the signal efficiencies of the identification and isolation requirements, as well as the number of events in sixteen categories, defined by whether each candidate passes or fails the photon identification and isolation criteria (four categories for each photon).
The signal leakage in the background control regions---those that fail the photon identification or isolation requirements---is evaluated with simulation.
A system of equations predicting the number of events in each category is solved with a $\chi^2$ minimisation process, extracting the background decomposition estimate shown in Table~\ref{tab:event_yields}.
The diphoton purity increases with the invariant mass $m_{\gamma\gamma}$ and varies from \SI{50}{\%} to \SI{90}{\%} across the mass range \SIrange{62}{120}{\GeV} with an uncertainty of \SI{1}{\%} to \SI{5}{\%} depending on the category.
The uncertainties in this purity measurement arise from: statistical uncertainty in the data and simulated MC samples; the alternate definitions of the identification criteria that define the control regions; the dependence of the signal leakage evaluation in the control regions on the event generator; the modelling of the isolation variable and shower-shape distributions; and possible correlations between the isolation variables and the inverted identification criteria~\cite{STDM-2012-05}.

\begin{table}
\centering
\caption{
The number of data events ($N_\textrm{data}$), the expected fraction of \gamgam, \gamjet, \jetjet events determined with the two dimensional side band method, and the fraction of Drell-Yan events in each category.
The uncertainties in the fractions of \gamgam, \gamjet, \jetjet arise from the statistical uncertainty varying the identification requirements.
The BDT categories are defined as follows: events with category BDT scores below $-0.2$ are in BDT Category 1, events with category BDT scores between $-0.2$ and 0 are in BDT Category 2, and events with category BDT scores above 0 are in BDT Category 3.
}
\renewcommand{\arraystretch}{1.15}
\begin{tabular}{cccccccc}
\toprule
Category                                  &                                 & UU              &  UC             & CC \\
\midrule
\multirow{4}{*}{Model Independent}        & $N_\textrm{data}$               & 1356130         & 1104590         & 243984    \\
& $f_{\gamgam}$ [\%]              & 74.4$\pm$1.3\phantom{0}            & 69.2$\pm$3.3            & 61.6$\pm$3.7 \\
& $f_{\gamjet}$ \phantom{.}[\%]   & 19.9$\pm$1.2\phantom{0}            & 23.7$\pm$1.9            & 28.1$\pm$2.2 \\
& $f_{\jetjet}$ \phantom{..}[\%]  & \phantom{0}5.3$\pm$0.5\phantom{0}  & \phantom{0}6.0$\pm$0.7  & \phantom{0}7.3$\pm$1.7 \\
& $f_\textrm{DY}$ [\%]            & 0.46$\pm$0.03  & \phantom{0}1.1$\pm$0.1  & \phantom{0}3.1$\pm$0.4  \\
\midrule
\multirow{4}{*}{BDT Category 1}           & $N_\textrm{data}$               & 592403          & 494228          & 112603   \\
& $f_{\gamgam}$ [\%]              & 71.5$\pm$2.0            & 67.0$\pm$3.4            & 57.3$\pm$3.8 \\
& $f_{\gamjet}$ \phantom{.}[\%]   & 20.9$\pm$1.5            & 24.1$\pm$2.0            & 29.9$\pm$3.1 \\
& $f_{\jetjet}$ \phantom{..}[\%]  & \phantom{0}6.8$\pm$0.6  & \phantom{0}7.2$\pm$0.9  & \phantom{0}8.2$\pm$0.9 \\
& $f_\textrm{DY}$ [\%]            & \phantom{0}0.7$\pm$0.1  & \phantom{0}1.7$\pm$0.2  & \phantom{0}4.7$\pm$0.6  \\
\midrule
\multirow{4}{*}{BDT Category 2}    & $N_\textrm{data}$                      & 508548          & 401646          & 86275 \\
& $f_{\gamgam}$ [\%]              & 74.7$\pm$1.4\phantom{0}            & 69.7$\pm$4.1\phantom{0}            & 64.5$\pm$3.8 \\
& $f_{\gamjet}$ \phantom{.}[\%]   & 20.2$\pm$1.3\phantom{0}            & 24.1$\pm$2.7\phantom{0}            & 26.7$\pm$2.6 \\
& $f_{\jetjet}$ \phantom{..}[\%]  & \phantom{0}4.8$\pm$0.5\phantom{0}  &  \phantom{0}5.5$\pm$0.6\phantom{0} & \phantom{0}7.0$\pm$1.2 \\
& $f_\textrm{DY}$ [\%]            & 0.29$\pm$0.02  & 0.69$\pm$0.01  & \phantom{0}1.8$\pm$0.2 \\
\midrule
\multirow{4}{*}{BDT Category 3}    & $N_\textrm{data}$                      & 255179          & 208716          & 45106   \\
& $f_{\gamgam}$ [\%]              & 80.4$\pm$1.0\phantom{0}            & 73.5$\pm$2.8\phantom{0}            & 66.6$\pm$5.1 \\
& $f_{\gamjet}$ \phantom{.}[\%]   & 16.7$\pm$1.2\phantom{0}            & 21.9$\pm$1.8\phantom{0}            & 26.2$\pm$3.3 \\
& $f_{\jetjet}$ \phantom{..}[\%]  & \phantom{0}2.7$\pm$0.3\phantom{0} & \phantom{0}4.0$\pm$0.9\phantom{0}  & \phantom{0}5.5$\pm$1.4 \\
& $f_\textrm{DY}$ [\%]            & 0.19$\pm$0.02  & 0.56$\pm$0.06  & \phantom{0}1.7$\pm$0.2  \\
\bottomrule
\end{tabular}
\label{tab:event_yields}
\end{table}

The continuum background \mgg distribution in each category is described by an analytic function whose form is determined using the method described in Ref.~\cite{HIGG-2016-21}.
The bias related to the choice of analytic function is estimated as the fitted \emph{spurious signal}~\cite{HIGG-2013-08} yield extracted using a signal-plus-background fit to a background-only template.
The background-only template is built using simulated samples for the \gamgam component and a data control sample for the \gamjet and \jetjet components, mixed according to the fractions determined by the 2x2D side band method presented above.
To minimise the effects of statistical fluctuations in the background-only template due to finite MC simulation statistics, the templates are smoothed with a Gaussian Process regression~\cite{GPML}.
This analysis uses a similar methodology to that used in Ref.~\cite{HIGG-2019-23}, with an update to use the Gibbs kernel~\cite{Gibbs:1997} to allow for an adaptive length scale hyperparameter.
The following functions are considered: Bernstein polynomials~\cite{Bernstein} of order five to seven, and exponentials of second, third and fourth order polynomials.
A function is only considered for modelling the background in a given analysis category if the maximal spurious signal from a fit to the background template is found to be less than 50\% of the statistical uncertainty in the fitted signal yield (from the background distributions normalised to the same statistics as the data) over the mass range \SIrange{66}{110}{\GeV}.
In the cases where two or more functions satisfy this requirement, the function with the fewest degrees of freedom is chosen to model the continuum background in that analysis category.
For most categories considered in this analysis, the result is an exponential of a third- or fourth-order polynomial.
The exception is the UC category in the model independent analysis, which is modelled using a Bernstein polynomial of order six.
The corresponding modelling uncertainty is derived by fitting the local maxima of the absolute spurious signal as a function of diphoton mass with an exponential of a second-order polynomial in each category.

The DY background is modelled using a DSCB function, with parameters determined by fitting a data-driven \mgg template.
At first order, this template is derived using an \mee distribution from $Z \rightarrow ee$ events in data.
Electrons are required to pass the same \ET and $\eta$ requirements as photons, as well as identification requirements based on shower shape variables; no isolation requirement is applied.
Because electrons misidentified as photons generally lose a large amount of energy due to bremsstrahlung, a Smirnov transformation~\cite{Devroye:1986:SNR:318242.318443} derived from simulation is used to correct the \mee template shape to the \mgg shape expected from electrons faking a photon.
The normalisation of the resulting \mgg template is computed from $e \rightarrow \gamma$ fake rates obtained from the same data used to derive the template, as described in Ref.~\cite{HIGG-2014-04}, and is shown in Table~\ref{tab:event_yields}.
Since electrons reconstructed as unconverted photons are more affected by bremsstrahlung than those reconstructed as converted photons, the UU events are more shifted to lower invariant masses than CC events.
To account for these differences, the Smirnov transformation and normalisation are derived separately for each analysis category.
Variations arising from limited MC sample sizes, variations of the $Z$-boson mass window and background subtraction used when deriving fake rates, results using different MC event generators, and results using different detector geometries are considered as uncertainties.


%
\section{Results}
\label{sec:results}
For the model-independent result, the measurement of the signal production cross-section times branching ratio, \xsecBR, is performed in a fiducial region.
The fiducial region, which closely matches the selection requirements of the reconstructed photons, is constructed in order to reduce the dependence of the measurement on the chosen theoretical model.
The fiducial region is defined at particle level as:
two photons with $\ET>\SI{22}{\GeV}$, $|\eta|<2.37$ excluding $1.37<|\eta|<1.52$,
passing the isolation requirement $\etiso<0.065\ \ET + \SI{1}{\GeV}$,
and passing $\ET/\mgg > 22/58$.
Here, $\etiso$ is defined as the magnitude of the vector sum of \pT for all particles with a lifetime longer than \SI{10}{\ps} (except neutrinos and muons) within a radius of $\Delta R = 0.2$ around the photon.
This isolation requirement is chosen to reproduce the detector-level selection.
The particle-level fiducial cross-section includes a signal efficiency correction factor \CX through:
\begin{equation}
\xsecBR = \frac{N_{\textrm{S}}}{\CX\mathcal{L}},\ \textrm{with}\
\CX = \frac{N^{\textrm{det}}_{\textrm{MC}}}{N^{\textrm{fid}}_{\textrm{MC}}},
\end{equation}
where $N_{\textrm{S}}$ is the fitted number of signal events in data,
$\mathcal{L}$ is the integrated luminosity,
$N^{\textrm{det}}_{\textrm{MC}}$ is the number of simulated signal events passing the detector-level selection criteria,
and $N^{\textrm{fid}}_{\textrm{MC}}$ is the number of simulated signal events passing the particle-level selection.
Since a generic spin-0 scalar $X$ is targeted, only the ggF production mode is used for the nominal correction factor, and the \CX factor values range from \SIrange{0.46}{0.62}{} as a function of \mX.
The envelope of the \CX values from the five production modes is taken as an uncertainty.
This model dependence uncertainty is determined using simulated samples of \X with several production modes: ggF, VBF, VH, and t\=tH.

Several experimental uncertainties directly impact the signal yields and signal shape and are accounted for in the fit using nuisance parameters constrained by Gaussian penalty terms in the likelihood function.
The largest effect on the signal yield in the model-dependent analysis comes from the uncertainty in the pile-up reweighting procedure.
This uncertainty is estimated by changing the nominal rescaling factor and evaluating the change in the signal efficiency.
Uncertainties in the signal yield arising from uncertainties in the luminosity determination, the trigger efficiency, and photon identification and isolation efficiencies are also considered.
Uncertainties in the signal mass scale and resolution are assessed by propagating the photon energy calibration uncertainties onto the MC signal samples and constructing a new signal template.
These shifted signal templates are then refit using the signal model parameterisation given in Section~\ref{sec:signal} and the deviations from the central mass and resolution values are taken as an uncertainty.
These uncertainties in the signal mass scale (resolution) are in the range $\pm\SI{0.3}{\%}$ to $\pm\SI{0.5}{\%}$ ($\pm\SI{3}{\%}$ to $\pm\SI{10}{\%}$), with the larger uncertainties occurring at higher values of \mgg.
A systematic uncertainty is derived from the variation of the detector material description in the simulation, resulting in migrations across conversion categories.
An uncertainty is also assessed on the electron-photon ambiguity BDT efficiency that accounts for differences in performance between data and the MC simulation selection efficiency in a control region outside of the mass range considered in this analysis.
The absolute differences in efficiencies between data and MC simulation are found to be \SI{0.1}{\%}, \SI{1.2}{\%}, and \SI{2.9}{\%} for the UU, UC, and CC categories, respectively.
Since no attempt was made to remove the \gamjet and \jetjet components from the diphoton data samples---accounting for around the \SI{25}{\%} of the data sample in the mass range \SIrange{110}{120}{\GeV}---these differences are expected to be conservative estimates.
The envelope of these uncertainties results in an uncertainty of \SI{0.7}{\%} on the signal selection efficiency.
The variations on the DY background result in uncertainties in the peak position, the peak width, and the normalisation which are derived separately for each conversion category.
The magnitude of these uncertainty components is shown in Table~\ref{tab:syst}.

For the model-dependent result, the total signal production cross-section times branching ratio, \xsecHBR, is computed as:
\begin{equation}
\xsecHBR = \frac{N_{\textrm{S}}}{\AH\CH\mathcal{L}},
\end{equation}
where $N_{S}$ is the number of signal events fit in data, \CH is the signal efficiency factor for a SM-like Higgs boson, and \AH is an acceptance factor defined as the probability of a generated event to be selected in the fiducial volume at the particle level.
The \AH and \CH factors are parameterised using the weighted average of the different Higgs boson production modes assuming the SM-like cross-section for each production mode.
The acceptance times correction factor varies from \SIrange{0.13}{0.2}{} depending on the mass of the SM-like Higgs boson \mH.
Since the model-dependent result assumes SM-like production-mode cross-sections when constructing \CH, the uncertainty from taking the envelope of the production modes---as done when deriving \CX in the model-independent results---is not relevant.
After accounting for the differences between the analyses, the model-dependent result provides an approximately \SI{30}{\%} to \SI{60}{\%} stronger limit than the model-independent result.

The expected fraction of signal events for the model-independent (model-dependent) analysis in each of the three conversion (nine) categories is parameterised as a function of \mX (\mH).
The migration of signal events between different BDT categories in the model-dependent analysis is found to be negligible.
The number of signal and background events is measured with an extended maximum-likelihood simultaneous binned fit to the \mgg spectra.
This fit is performed in the three conversion categories for the model-independent analysis and in nine categories (three BDT categories for each conversion category) in the model-dependent analysis.
Scans over different \mX and \mH hypotheses are performed in 0.1 \GeV\ steps over the mass range \SIrange{66}{110}{\GeV}.

The signal-plus-background model has the form:
\begin{equation}
N_{\textrm{S}} \cdot f_{\textrm{S}}(\mgg)
+ N_{\textrm{SS}} \cdot f_{\textrm{S}}(\mgg)
+ N_{\textrm{B}} \cdot f_{\textrm{B}}(\mgg) + N_{\textrm{DY}} \cdot f_{\textrm{DY}}(\mgg),
\end{equation}
where $f_{\textrm{S}}$ is the signal model described in Section~\ref{sec:signal}, $N_{\textrm{S}}$ ($N_{\textrm{B}}$) is the fitted number
of signal (continuum background) events, $N_{\textrm{SS}}$ is the number of spurious signal events, $f_{\textrm{B}}$ ($f_{\textrm{DY}}$) is the
continuum (Drell-Yan) background model described in Section~\ref{sec:background}, and $N_\textrm{DY}$ is the number of DY background events.
The systematic uncertainties are included in the
likelihood via nuisance parameters, and constrained by Gaussian or log-normal penalty terms.

\begin{figure}
\centering
\subfloat[]{
\includegraphics[width=0.5\textwidth]{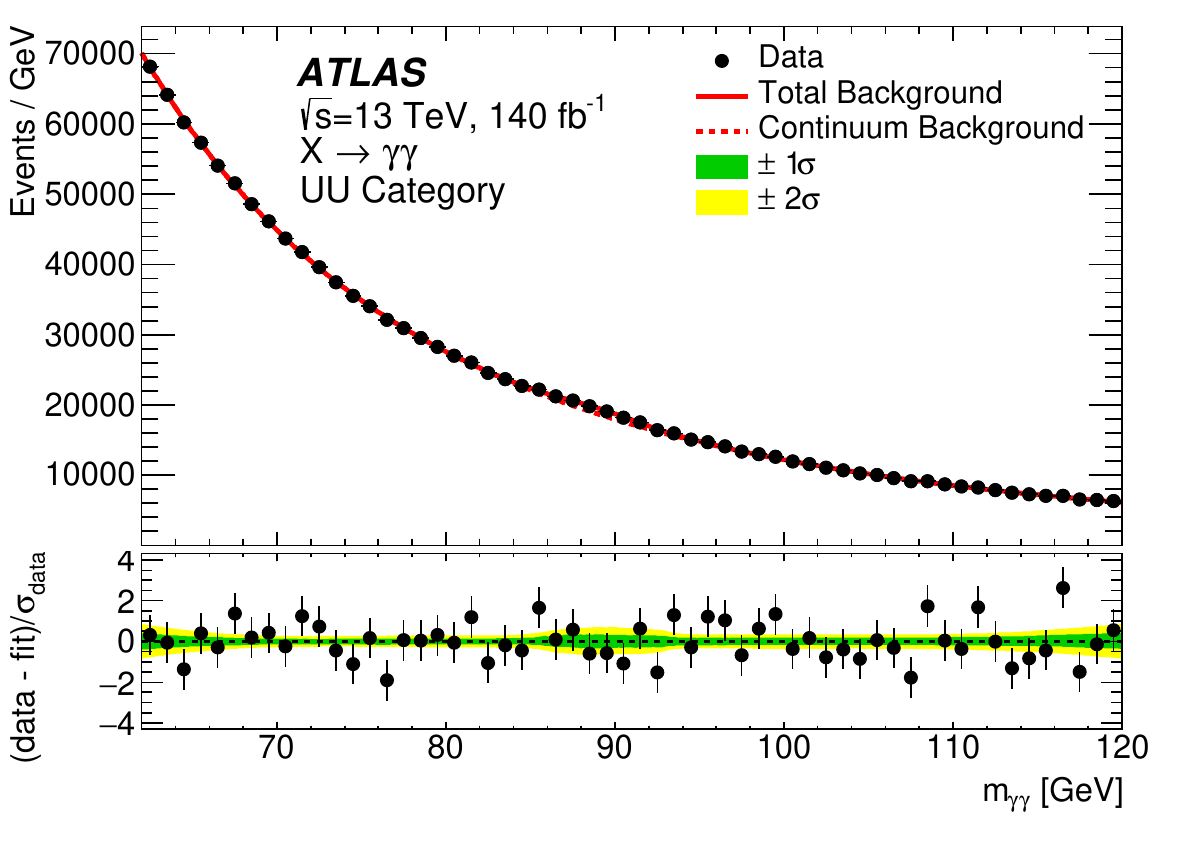}
}
\subfloat[]{
\includegraphics[width=0.5\textwidth]{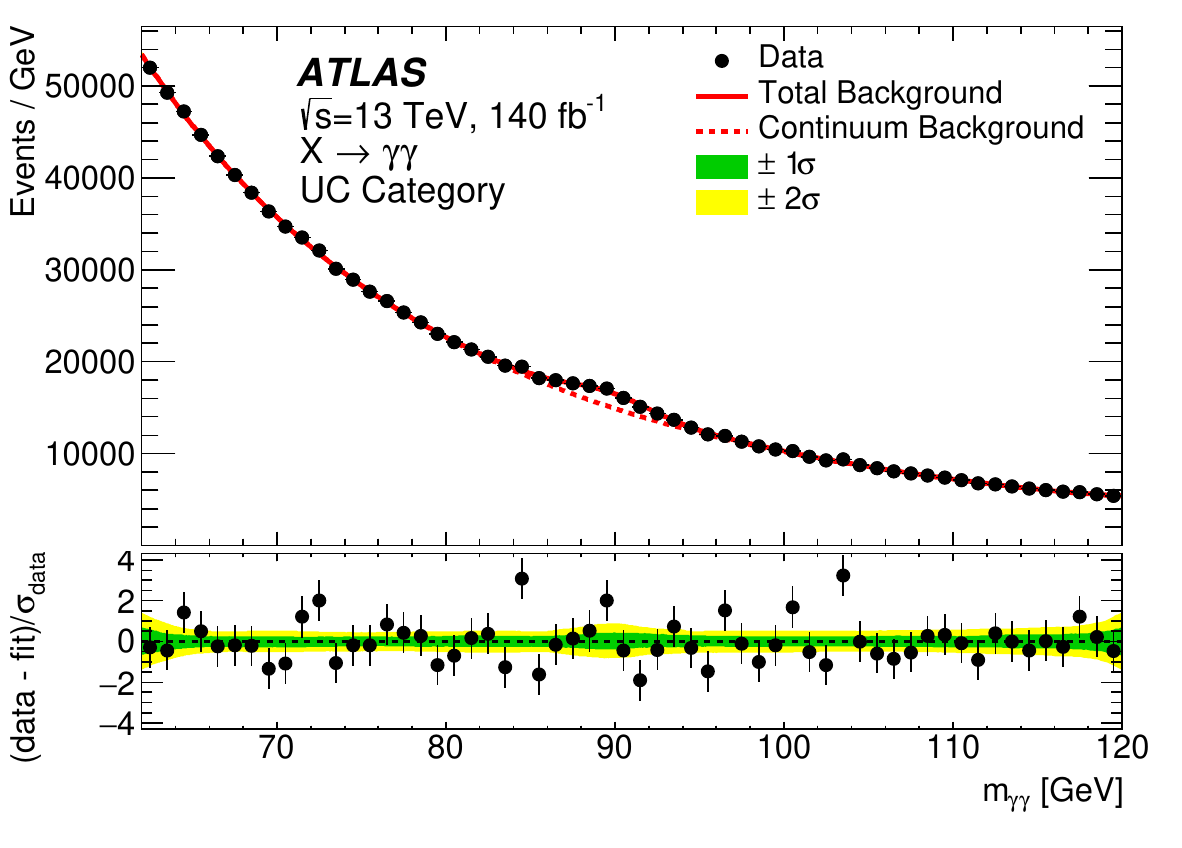}
}\\
\subfloat[]{
\includegraphics[width=0.5\textwidth]{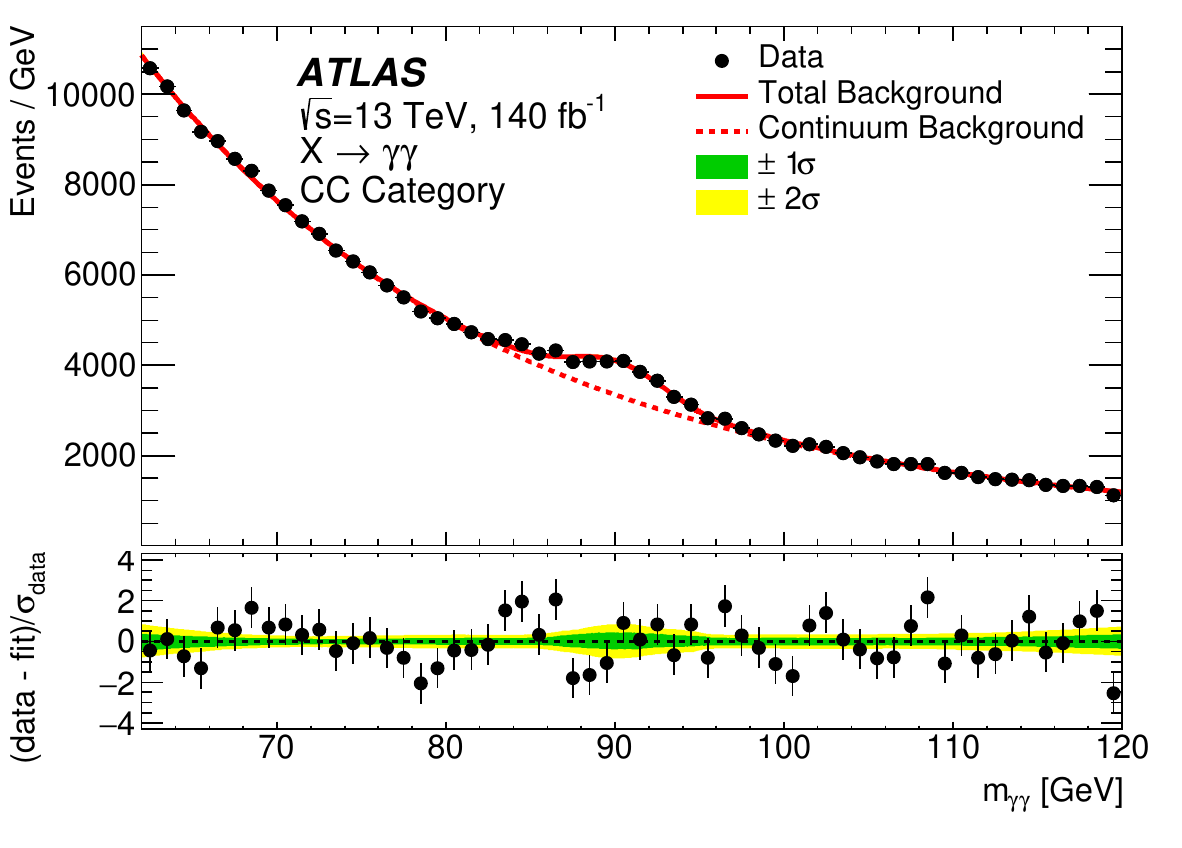}
}
\caption{
Background-only fit to the data (black markers) as a function of the diphoton invariant mass \mgg for the model-independent conversion categories: (a) UU, (b) UC, and (c) CC.
The solid lines show the sum of the Drell-Yan and the continuum background components and the dashed lines show only the continuum background components.
The difference between the data and the total background component divided by the statistical uncertainty of the data, $\sigma_\mathrm{data}$, is shown at the bottom panel separately for each category.
The green (yellow) bands denote the total uncertainty in the background model at one (two) standard deviation.
}
\label{fig:background-fit-inc}
\end{figure}

\begin{figure}
\centering
\subfloat[]{
\includegraphics[width=0.5\textwidth]{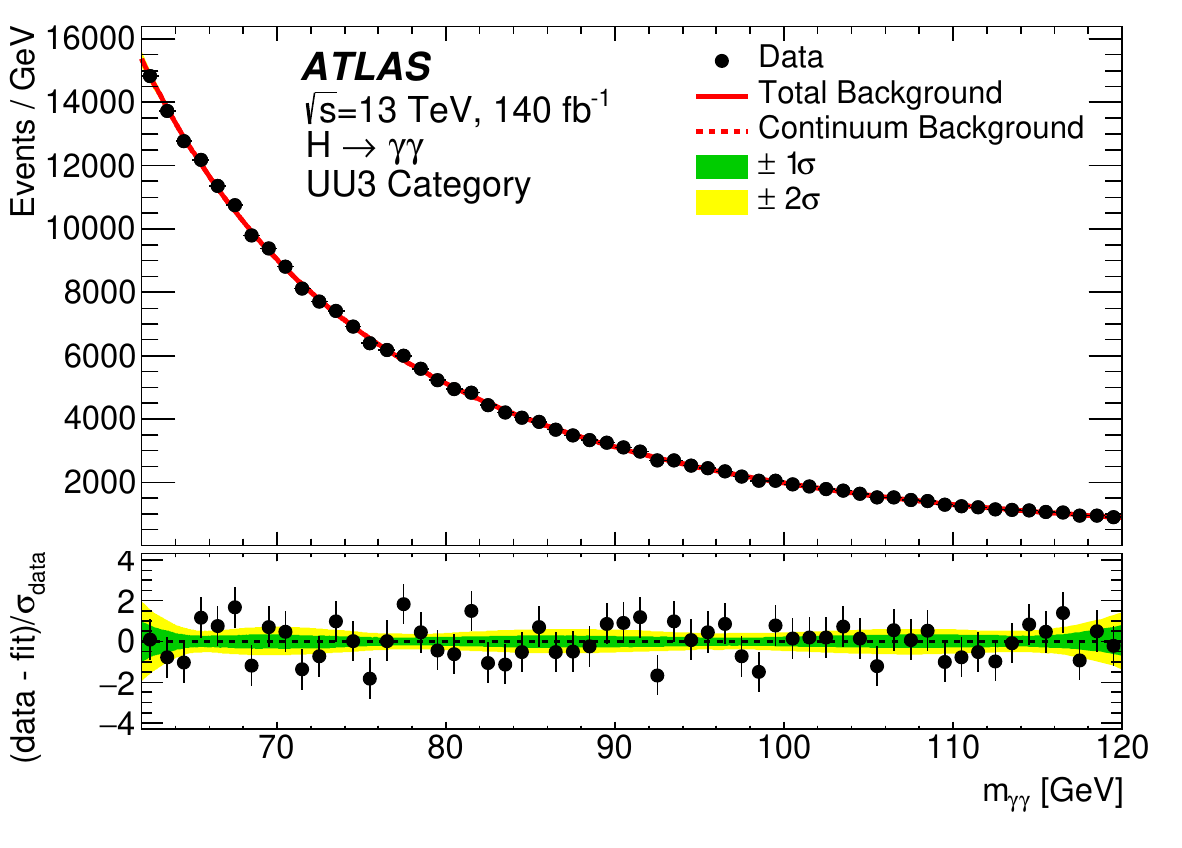}
}
\subfloat[]{
\includegraphics[width=0.5\textwidth]{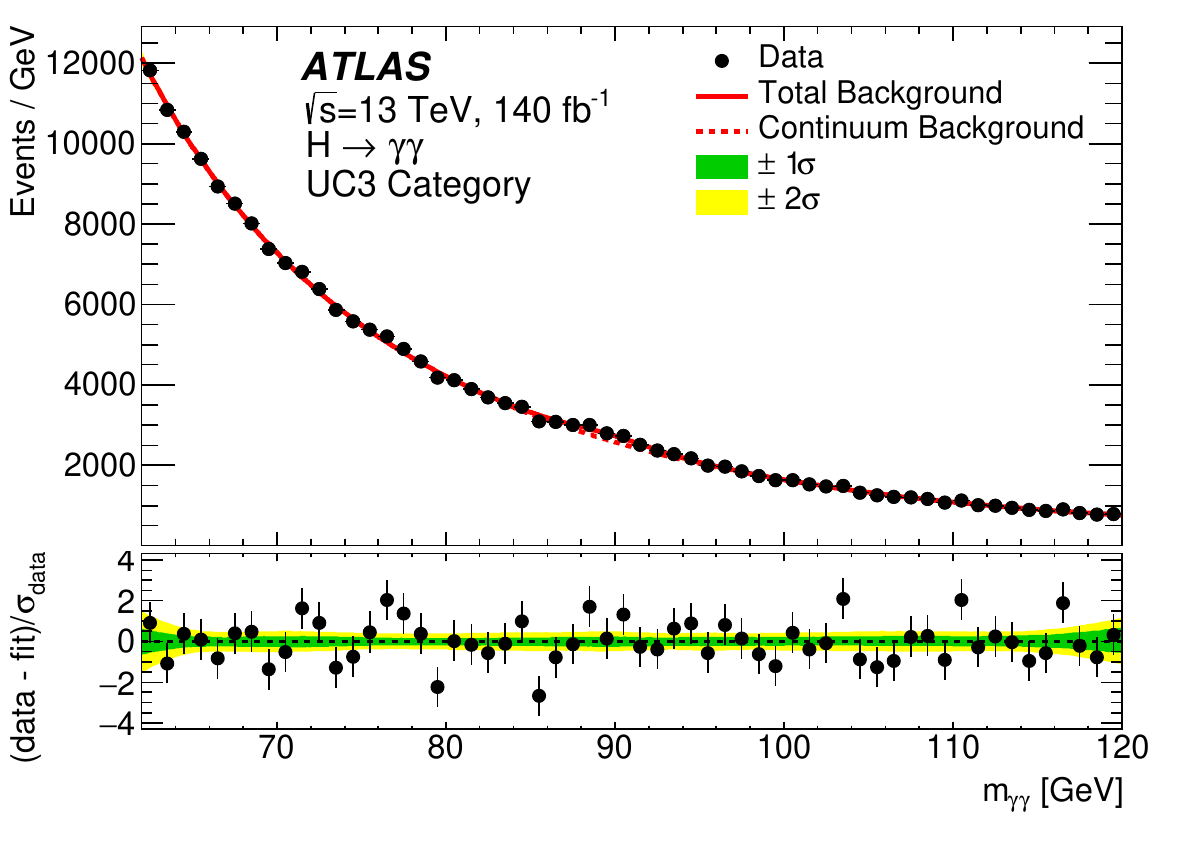}
}\\
\subfloat[]{
\includegraphics[width=0.5\textwidth]{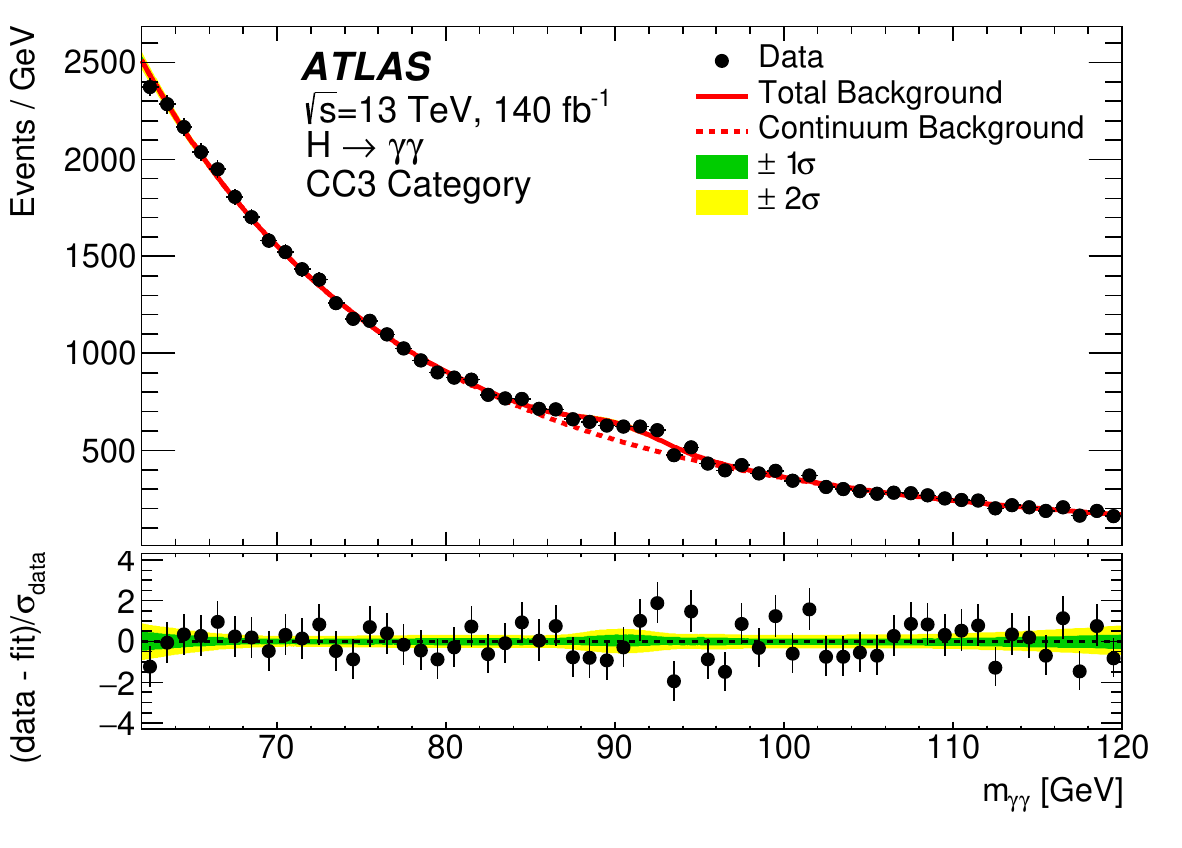}
}
\caption{
Background-only fit to the data (black markers) as a function of the diphoton invariant mass \mgg for each of the model-dependent BDT 3 categories: (a) UU3, (b) UC3, and (c) CC3.
The solid lines show the sum of the Drell-Yan and the continuum  background components and the dashed lines show only the continuum background components.
The difference between the data and the total background component divided by the uncertainty, with $\sigma_\mathrm{data}$ denoting only the statistical error of the data, is shown at the bottom panel separately for each category.
The green (yellow) bands denote the total uncertainty in the background model at one (two) standard deviation.
}
\label{fig:background-fit-bdt3}
\end{figure}

A background-only fit of the data is shown in Figure~\ref{fig:background-fit-inc} for the three model-independent conversion categories and in Figure~\ref{fig:background-fit-bdt3} for the three model-dependent BDT 3 categories.
The parameters of the background model in each category are determined during a simultaneous fit to all categories.
A goodness-of-fit test is performed in each category and returns a probability between 5\% and 89\% depending on the category.
As expected, the DY contribution is most prominent in the CC categories (see Table~\ref{tab:event_yields}).

The compatibility of the observed diphoton mass spectra with the background-only
hypothesis, for a given resonance mass, is determined with a local
$p$-value based on the profile-likelihood-ratio-test statistic~\cite{Cowan:2010js} as detailed in Ref.~\cite{HIGG-2016-08}.
The observed and expected \SI{95}{\%} confidence level (CL) exclusion limits on the production cross-section
times branching ratio are evaluated
using a modified frequentist approach $\textrm{CL}_\textrm{s}$~\cite{Read:2002hq} with the asymptotic
approximation to the test-statistic distribution~\cite{Cowan:2010js}.

\begin{table}[tb]
\caption{Summary of the systematic uncertainties considered in this analysis.
In general, the values correspond to the uncertainties associated to the fit nuisance parameters.
The DY uncertainty is the percent error on the nominal peak position, peak width, and normalisation.
The spurious signal uncertainty is expressed as a number of events and relative to the expected statistical uncertainty ($\delta S$) of a fitted signal.
The ``Remarks'' column indicates specific information about the systematic uncertainty, including whether or not the uncertainty varies as a function of resonance mass or analysis category.
}
\label{tab:syst}
\begin{center}
\begin{footnotesize}
\begin{tabular}{@{}lcl@{}}
\toprule
Source            &  Uncertainty [\%] & Remarks \\
\midrule
\multicolumn{3}{l}{\em Signal yield} \\
\hspace{.4cm} Luminosity                       & $\pm 0.83$  &  \\
\hspace{.4cm} Electron-photon ambiguity BDT efficiency & $\pm 0.7$ &  \\
\hspace{.4cm} Trigger efficiency               & $\pm 1.0$\,\text{--}\,1.5  & \mX-dependent \\
\hspace{.4cm} Photon identification efficiency & $\pm 1.8$\,\text{--}\,3.0  & \mX-dependent \\
\hspace{.4cm} Photon isolation efficiency      & $\pm 1.6$\,\text{--}\,2.4  & \mX-dependent \\
\hspace{.4cm} Photon energy scale              & $\pm 0.1$\,\text{--}\,0.3  & \mX-dependent \\
\hspace{.4cm} Photon energy resolution         & $\pm 0.1$\,\text{--}\,$0.15$  & \mX-dependent \\
\hspace{.4cm} Pile-up                          & $\pm 1.6$\,\text{--}\,5.0  & \mX-dependent \\
\hspace{.4cm} Production mode                  & $\pm 4.3$\,\text{--}\,29   & \mX-dependent (model-independent only) \\
\midrule
\multicolumn{3}{l}{\em Signal modelling} \\
\hspace{.4cm} Photon energy scale        & $\pm 0.3$\,\text{--}\,0.5 & \mX-\ and category--dependent  \\
\hspace{.4cm} Photon energy resolution   & $\pm 3$\,\text{--}\,10 & \mX-\ and category-dependent  \\
\midrule
\multicolumn{3}{l}{\em Migration between categories} \\
\hspace{.4cm} Material                   & $-2.0$\,/\,$+1.0$\,/\,$+4.1$ & category-dependent \\
\midrule
\multicolumn{3}{l}{\em DY background modelling} \\
\hspace{.4cm} Peak position              & $\pm 0.1$\,\text{--}\,$0.2$ & category-dependent   \\
\hspace{.4cm} Peak width                 & $\pm 1.9$\,\text{--}\,$3.5$ & category-dependent   \\
\hspace{.4cm} Normalisation              & $\pm 7.1$\,\text{--}\,$13$ & category-dependent   \\
\midrule
\multicolumn{3}{l}{\em Continuum background (model-dependent) } \\
\hspace{.4cm} Spurious signal, NWA       & $9$\,\text{--}\,$171$ events, (10\%\,\text{--}\,50\%) & \mX-\ and category-dependent \\
\multicolumn{3}{l}{\em Continuum background (model-independent) } \\
\hspace{.4cm} Spurious signal, NWA       & $37$\,\text{--}\,$310$ events, (20\%\,\text{--}\,50\%) & \mX-\ and category-dependent \\
\hspace{.4cm} Spurious signal, $\Gamma_{X}/\mX = 1.0\%$ & $65$\,\text{--}\,$539$ events, (20\%\,\text{--}\,50\%) & \mX-\ and category-dependent \\
\hspace{.4cm} Spurious signal, $\Gamma_{X}/\mX = 2.5\%$ & $92$\,\text{--}\,$879$ events, (20\%\,\text{--}\,50\%) & \mX-\ and category-dependent \\
\bottomrule
\end{tabular}
\end{footnotesize}
\end{center}
\end{table}

\begin{figure}[htbp]
\centering
\subfloat[]{
\includegraphics[width=0.5\columnwidth]{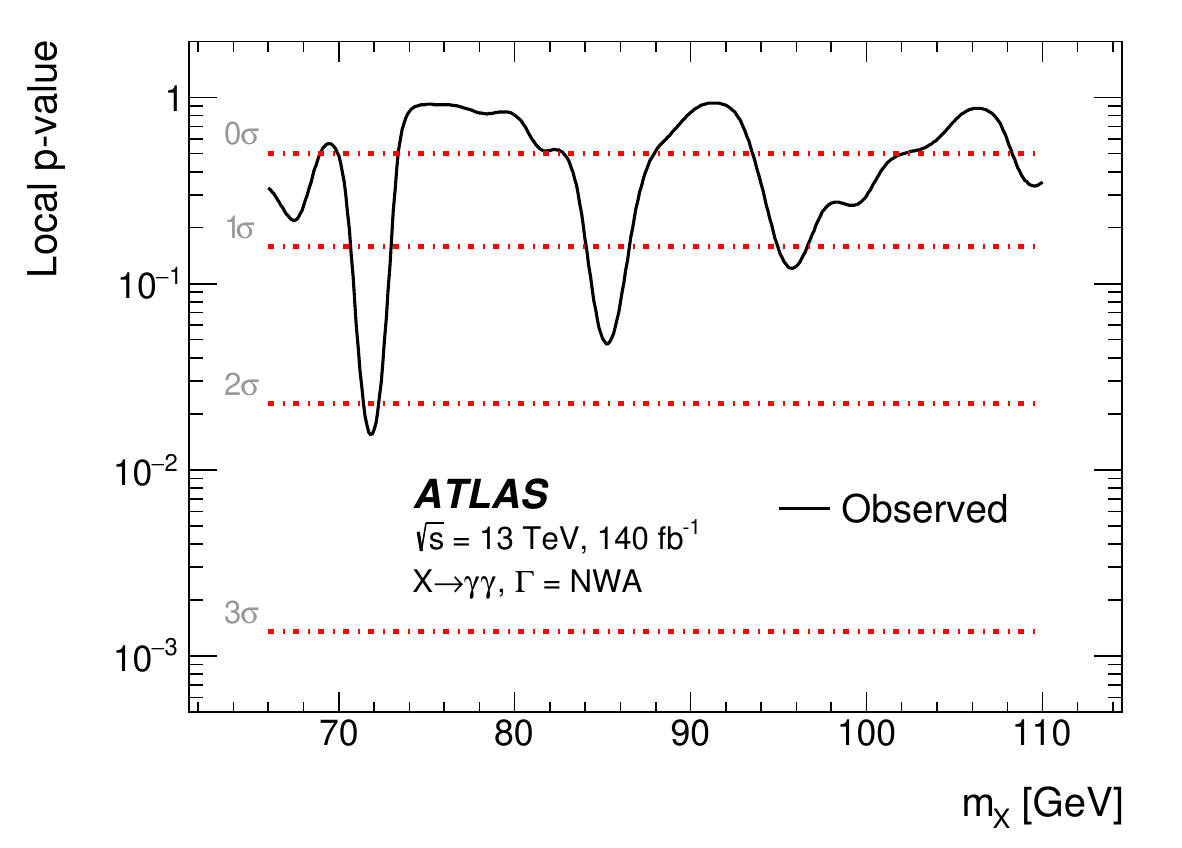}
\label{fig:p0-indep}
}
\subfloat[]{
\includegraphics[width=0.5\columnwidth]{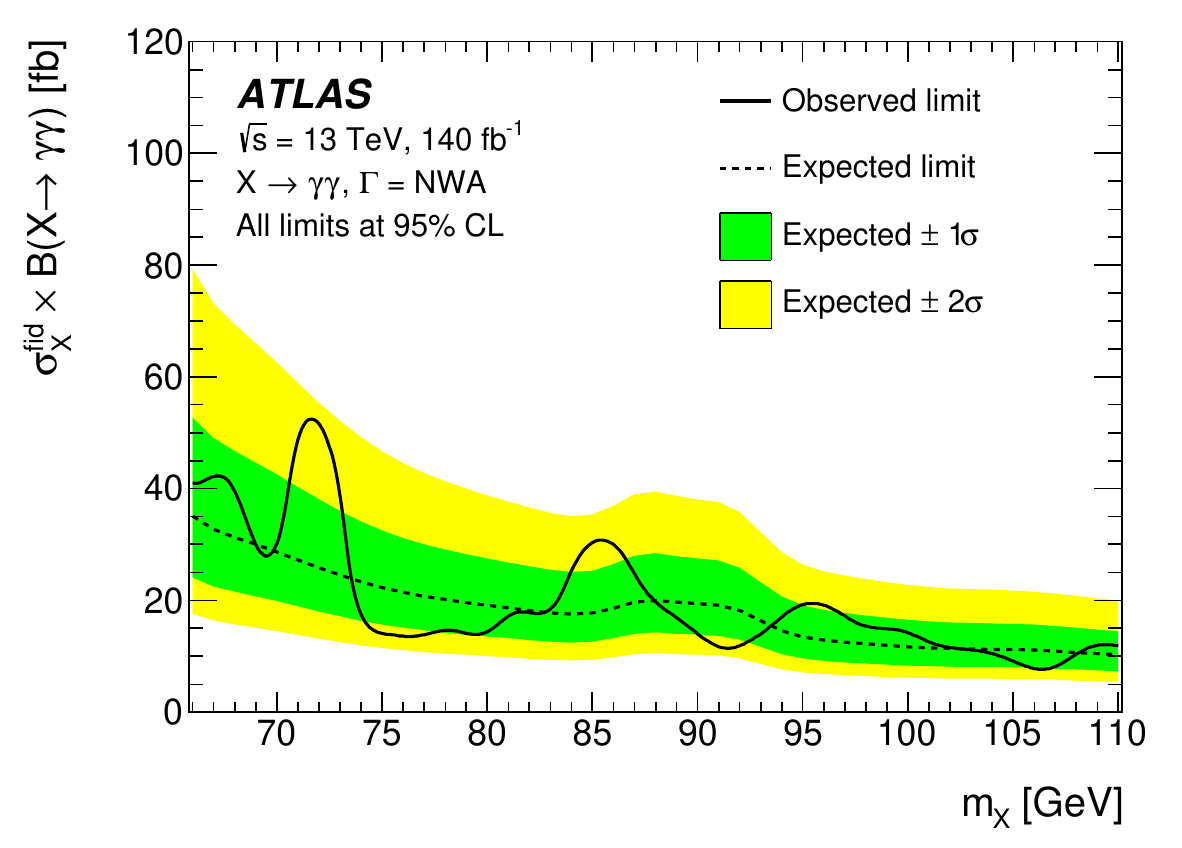}
\label{fig:limit-indep}
}
\caption{(a) Compatibility of the data, in the model-independent search, in terms of local $p$-value (solid line), with the
background-only hypothesis as a function of the assumed NWA signal mass \mX.
The dotted-dashed lines correspond to the standard deviation quantification $\sigma$.
(b) 95\% CL upper limits on the fiducial cross-section times branching ratio \BRxgg as a
function of NWA \mX, where the solid (dashed) line corresponds to the observed (expected) limit and
the green (yellow) band corresponds to one (two) standard deviation from the expectation.}
\end{figure}

The result of the model-independent $p$-value scan is shown in Figure~\ref{fig:p0-indep}.
No significant excess with respect to the background-only hypothesis is observed.
The largest localised deviation for the model-independent search is observed for a mass of 71.8~\GeV, corresponding to a local significance of 2.2$\sigma$.

For the narrow-width model-independent result, an upper limit at the \SI{95}{\%} CL is set on \xsecBR from 8\,fb to 53\,fb, as shown in Figure~\ref{fig:limit-indep}.

\begin{figure}[bp]
\centering
\includegraphics[width=0.5\columnwidth]{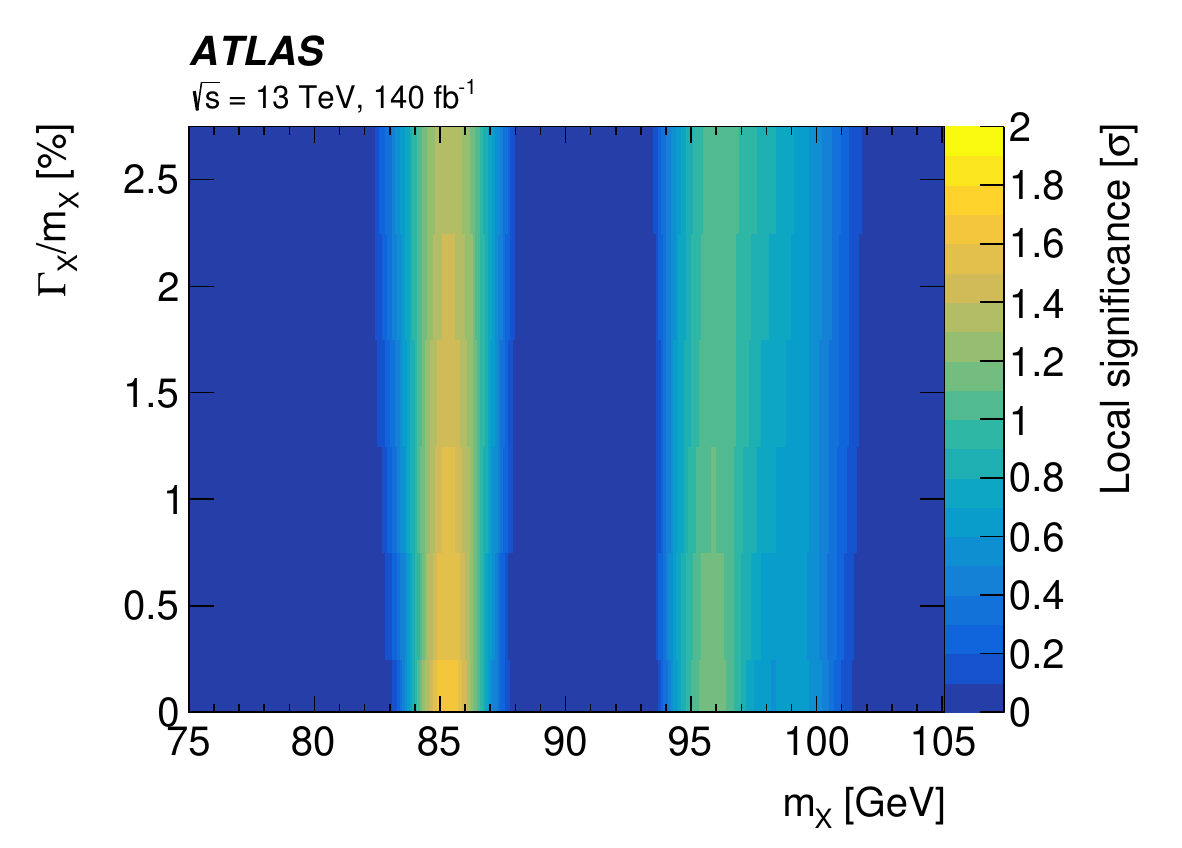}
\caption{Compatibility of the data with the background-only hypothesis, using the local $p_{0}$ quantified in units of standard deviations, $\sigma$, as a function of the assumed signal mass $m_{\textrm{X}}$ and of the relative width $\Gamma_{\textrm{X}}$/$m_{\textrm{X}}$ for the model-independent search.
}
\label{fig:2D_significance}
\end{figure}

\begin{figure}[tbp]
\centering
\subfloat[]{
\includegraphics[width=0.5\columnwidth]{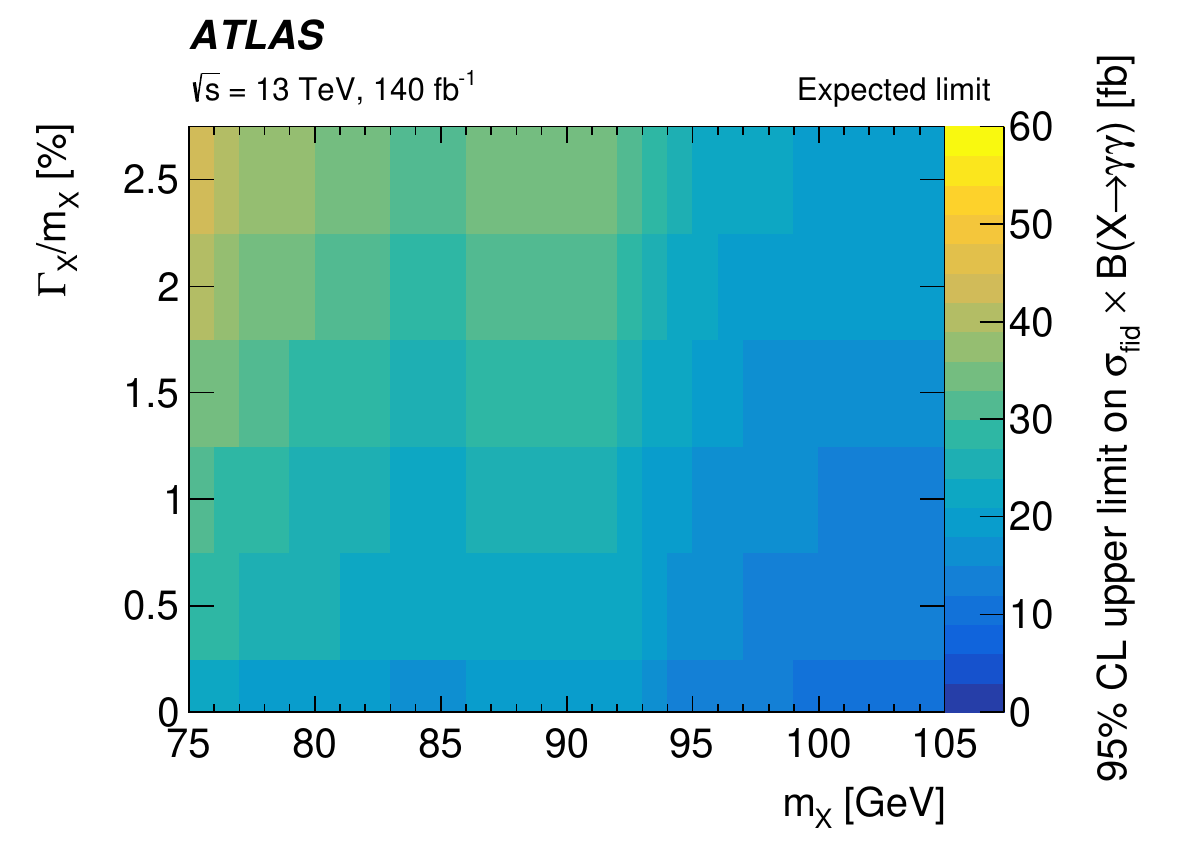}
}
\subfloat[]{
\includegraphics[width=0.5\columnwidth]{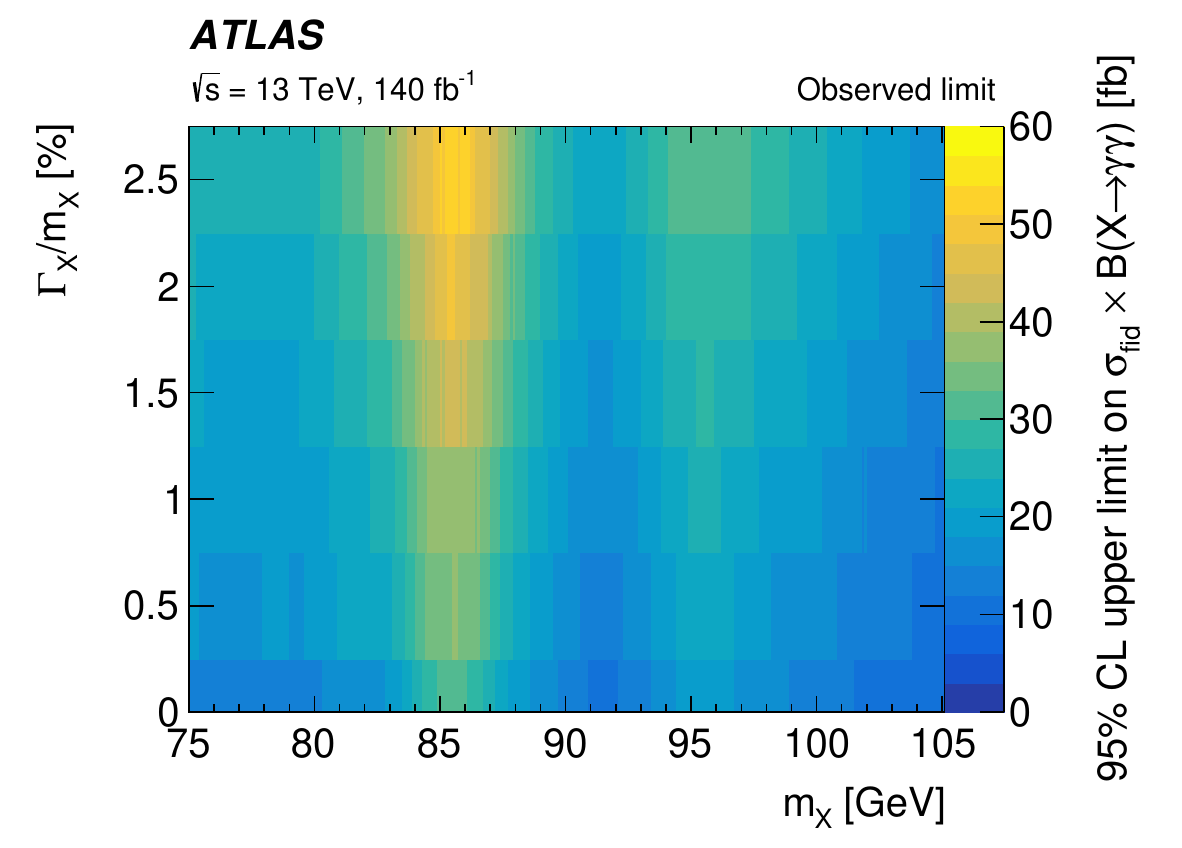}
}
\caption{
(a) Expected and (b) observed limits on the fiducial cross-section times branching ratio \BRxgg computed using the asymptotic approximation as a function of the assumed signal mass $m_{\textrm{X}}$ and relative width $\Gamma_{\textrm{X}}/m_{\textrm{X}}$ for the model-independent scalar resonance search.}
\label{fig:2D_limits}
\end{figure}

The model-independent analysis considers additional larger width signal hypotheses, and the $p$-value scan is expanded to include signal hypotheses with $\Gamma_{\textrm{X}}/m_{\textrm{X}}$ in the range \SI{0}{\%} to \SI{2.5}{\%}.
The result of this $p$-value scan is shown in Figure~\ref{fig:2D_significance}.
Due to the truncated range in $m_{\textrm{X}}$, the most significant excess occurs at \SI{85.2}{\GeV} for the narrow-width model hypothesis and corresponds to a local significance of $1.7\sigma$.
In the absence of a significant excess, upper limits are set on \xsecBR as a function of $\Gamma_{\textrm{X}}/m_{\textrm{X}}$ and are illustrated in Figure~\ref{fig:2D_limits}.

\begin{figure}[bp]
\centering
\subfloat[]{
\includegraphics[width=0.5\columnwidth]{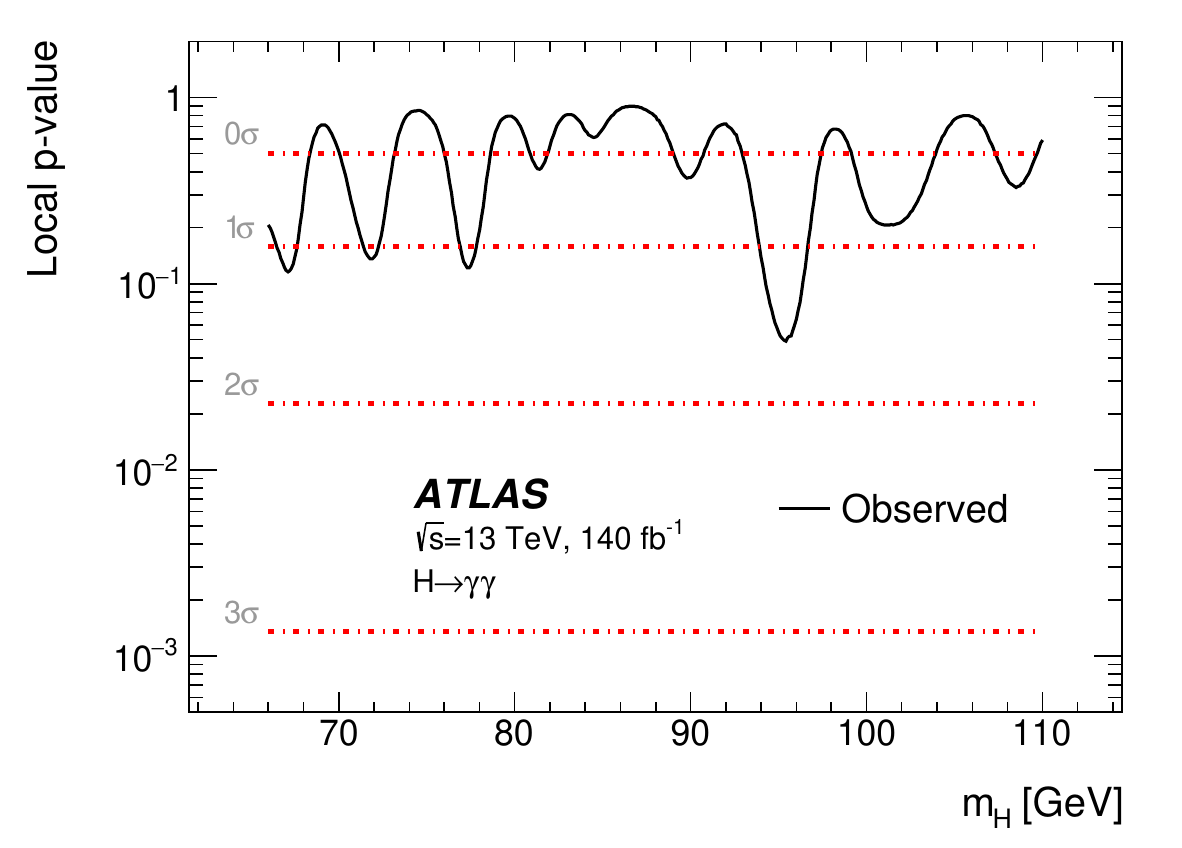}
\label{fig:p0-dep}
}
\subfloat[]{
\includegraphics[width=0.5\columnwidth]{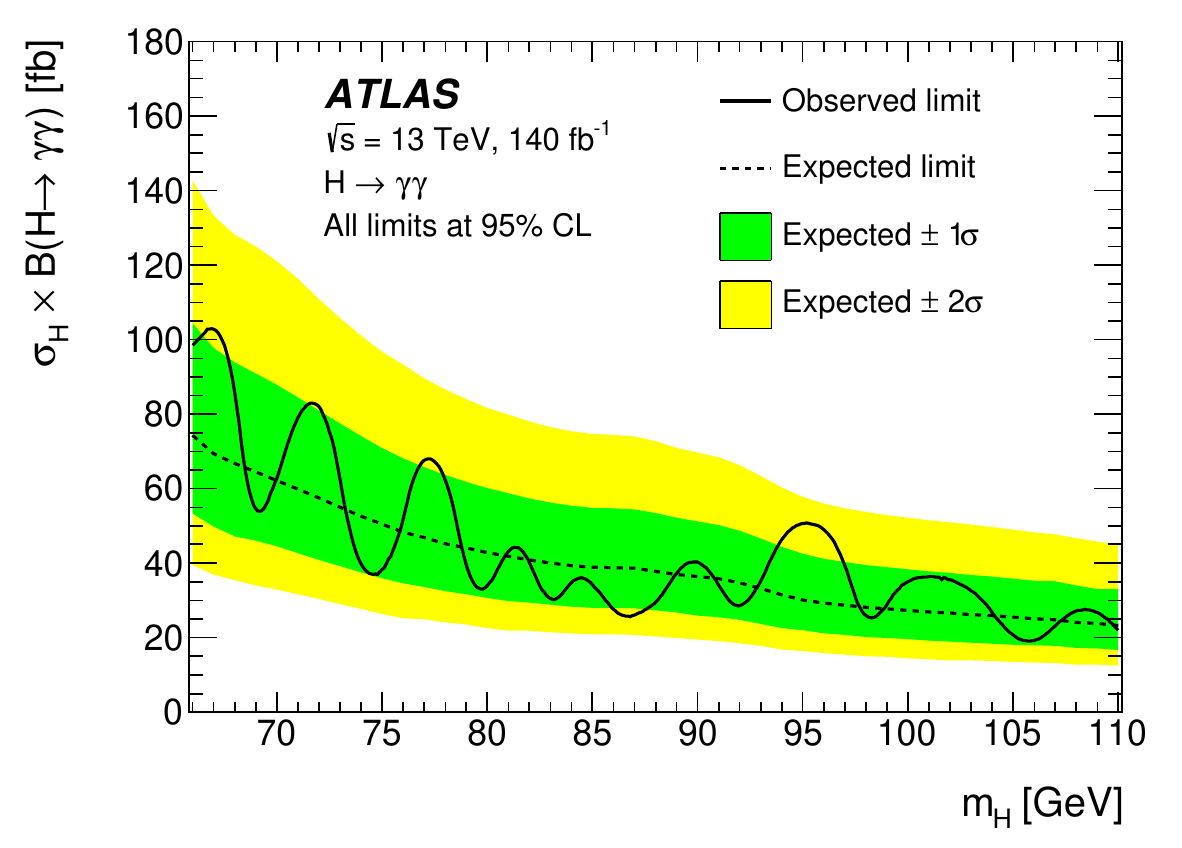}
\label{fig:limit-dep}
}
\caption{(a) Compatibility of the data with the background-only hypothesis as quantified by the local $p$-value (solid line) as a function of the assumed signal mass \mH, for the model-dependent search.
The dotted-dashed lines correspond to the standard deviation quantification $\sigma$.
(b) 95\% CL upper limits on the total cross-section times branching ratio \BRhgg as a
function of \mH, where the solid (dashed) line corresponds to the observed (expected) limit and
the green (yellow) band corresponds to one (two) standard deviation from the expectation.}
\end{figure}

The result of the model-dependent $p$-value scan is shown in Figure~\ref{fig:p0-dep} and the largest deviation is observed for a mass of \SI{95.4}{GeV}, corresponding to a local significance of 1.7$\sigma$.
An upper limit at the \SI{95}{\%} CL is set on \xsecHBR from \SI{19}{\fb} to \SI{100}{\fb} for the model-dependent result, as shown in Figure~\ref{fig:limit-dep}.
The limited number of $pp$ collisions recorded is the dominant uncertainty impacting this result.
The model-dependent result can be compared to a similar result from the CMS Collaboration, which sets an observed upper limit ranging from \SI{15}{\fb} to \SI{73}{\fb} in the mass range \SIrange{70}{110}{\GeV}~\cite{CMS-HIG-20-002}.
The largest deviation observed by CMS is for a mass of \SI{95.4}{\GeV}, corresponding to a local significance of 2.9$\sigma$.

Despite using the same $pp$ collision events, the results between the model-dependent and model-independent searches in this analysis are expected to show differences due to the model assumptions and additional categories used in the model-dependent analysis.
An example that illustrates these differences is the mild excess of events that appears around \mH = \SI{77}{\GeV} in the model-dependent analysis.
This mild excess is localised in two high-significance categories, UU3 and UC3, which contain about \SI{20}{\%} of all UU and UC events.
In the low-significance categories that comprise the bulk of the UU and UC events, a deficit of events is observed, which matches that observed in the model-independent analysis.

\FloatBarrier


\FloatBarrier

%
\section{Conclusion}
Searches for new narrow-width resonances, either a generic spin-0 particle or an additional low-mass Higgs boson, are performed in the diphoton invariant mass spectra ranging from 66 \GeV\ to 110 \GeV, using \SI{140}{\per\fb} of $pp$ collision data collected at $\sqrt{s}=\SI{13}{\TeV}$ with the ATLAS detector at the Large Hadron Collider.
The dominant uncertainties arise from the limited number of $pp$ collisions collected and the spurious signal uncertainty due to the choice of analytic functions to model the continuum background.
Both a model-independent search for a spin-0 particle ($X$) and a model-dependent search for an additional low-mass Higgs boson ($H$, assuming the SM production-mode times branching ratio to two photons cross-sections) are performed.
No significant excess above the SM background expectation is observed, and \SI{95}{\%} CL upper limits on the cross-section times branching ratio are set for each search.
In the model-independent analysis, the observed \SI{95}{\%} CL upper limits on the fiducial cross-section times branching ratio for a generic spin-0 signal are in the range \SI{8}{\fb} to \SI{53}{\fb} for new resonances with masses $66 < \mX < \SI{110}{\GeV}$.
For the model-dependent analysis, the observed upper limits on the production cross-section times branching ratio to two photons for a SM-like Higgs boson range from \SI{19}{\fb} to \SI{102}{\fb} in the same mass range.
%
%
%


%
\section*{Acknowledgements}
%

%
%

%
%

We thank CERN for the very successful operation of the LHC and its injectors, as well as the support staff at
CERN and at our institutions worldwide without whom ATLAS could not be operated efficiently.

The crucial computing support from all WLCG partners is acknowledged gratefully, in particular from CERN, the ATLAS Tier-1 facilities at TRIUMF/SFU (Canada), NDGF (Denmark, Norway, Sweden), CC-IN2P3 (France), KIT/GridKA (Germany), INFN-CNAF (Italy), NL-T1 (Netherlands), PIC (Spain), RAL (UK) and BNL (USA), the Tier-2 facilities worldwide and large non-WLCG resource providers. Major contributors of computing resources are listed in Ref.~\cite{ATL-SOFT-PUB-2023-001}.

We gratefully acknowledge the support of ANPCyT, Argentina; YerPhI, Armenia; ARC, Australia; BMWFW and FWF, Austria; ANAS, Azerbaijan; CNPq and FAPESP, Brazil; NSERC, NRC and CFI, Canada; CERN; ANID, Chile; CAS, MOST and NSFC, China; Minciencias, Colombia; MEYS CR, Czech Republic; DNRF and DNSRC, Denmark; IN2P3-CNRS and CEA-DRF/IRFU, France; SRNSFG, Georgia; BMBF, HGF and MPG, Germany; GSRI, Greece; RGC and Hong Kong SAR, China; ISF and Benoziyo Center, Israel; INFN, Italy; MEXT and JSPS, Japan; CNRST, Morocco; NWO, Netherlands; RCN, Norway; MEiN, Poland; FCT, Portugal; MNE/IFA, Romania; MESTD, Serbia; MSSR, Slovakia; ARIS and MVZI, Slovenia; DSI/NRF, South Africa; MICINN, Spain; SRC and Wallenberg Foundation, Sweden; SERI, SNSF and Cantons of Bern and Geneva, Switzerland; NSTC, Taipei; TENMAK, T\"urkiye; STFC/UKRI, United Kingdom; DOE and NSF, United States of America.

Individual groups and members have received support from BCKDF, CANARIE, CRC and DRAC, Canada; CERN-CZ, PRIMUS 21/SCI/017 and UNCE SCI/013, Czech Republic; COST, ERC, ERDF, Horizon 2020, ICSC-NextGenerationEU and Marie Sk{\l}odowska-Curie Actions, European Union; Investissements d'Avenir Labex, Investissements d'Avenir Idex and ANR, France; DFG and AvH Foundation, Germany; Herakleitos, Thales and Aristeia programmes co-financed by EU-ESF and the Greek NSRF, Greece; BSF-NSF and MINERVA, Israel; Norwegian Financial Mechanism 2014-2021, Norway; NCN and NAWA, Poland; La Caixa Banking Foundation, CERCA Programme Generalitat de Catalunya and PROMETEO and GenT Programmes Generalitat Valenciana, Spain; G\"{o}ran Gustafssons Stiftelse, Sweden; The Royal Society and Leverhulme Trust, United Kingdom.

In addition, individual members wish to acknowledge support from CERN: European Organization for Nuclear Research (CERN PJAS); Chile: Agencia Nacional de Investigaci\'on y Desarrollo (FONDECYT 1190886, FONDECYT 1210400, FONDECYT 1230812, FONDECYT 1230987); China: National Natural Science Foundation of China (NSFC - 12175119, NSFC 12275265, NSFC-12075060); Czech Republic: PRIMUS Research Programme (PRIMUS/21/SCI/017); EU: H2020 European Research Council (ERC - 101002463); European Union: European Research Council (ERC - 948254), Horizon 2020 Framework Programme (MUCCA - CHIST-ERA-19-XAI-00), European Union, Future Artificial Intelligence Research (FAIR-NextGenerationEU PE00000013), Italian Center for High Performance Computing, Big Data and Quantum Computing (ICSC, NextGenerationEU), Marie Sklodowska-Curie Actions (EU H2020 MSC IF GRANT NO 101033496); France: Agence Nationale de la Recherche (ANR-20-CE31-0013, ANR-21-CE31-0013, ANR-21-CE31-0022, ANR-22-EDIR-0002), Investissements d'Avenir Idex (ANR-11-LABX-0012), Investissements d'Avenir Labex (ANR-11-LABX-0012); Germany: Baden-Württemberg Stiftung (BW Stiftung-Postdoc Eliteprogramme), Deutsche Forschungsgemeinschaft (DFG - 469666862, DFG - CR 312/5-1); Italy: Istituto Nazionale di Fisica Nucleare (FELLINI G.A. n. 754496, ICSC, NextGenerationEU); Japan: Japan Society for the Promotion of Science (JSPS KAKENHI JP21H05085, JSPS KAKENHI JP22H01227, JSPS KAKENHI JP22H04944, JSPS KAKENHI JP22KK0227); Netherlands: Netherlands Organisation for Scientific Research (NWO Veni 2020 - VI.Veni.202.179); Norway: Research Council of Norway (RCN-314472); Poland: Polish National Agency for Academic Exchange (PPN/PPO/2020/1/00002/U/00001), Polish National Science Centre (NCN 2021/42/E/ST2/00350, NCN OPUS nr 2022/47/B/ST2/03059, NCN UMO-2019/34/E/ST2/00393, UMO-2020/37/B/ST2/01043, UMO-2021/40/C/ST2/00187); Slovenia: Slovenian Research Agency (ARIS grant J1-3010); Spain: BBVA Foundation (LEO22-1-603), Generalitat Valenciana (Artemisa, FEDER, IDIFEDER/2018/048), La Caixa Banking Foundation (LCF/BQ/PI20/11760025), Ministry of Science and Innovation (MCIN \& NextGenEU PCI2022-135018-2, MICIN \& FEDER PID2021-125273NB, RYC2019-028510-I, RYC2020-030254-I, RYC2021-031273-I, RYC2022-038164-I), PROMETEO and GenT Programmes Generalitat Valenciana (CIDEGENT/2019/023, CIDEGENT/2019/027); Sweden: Swedish Research Council (VR 2018-00482, VR 2022-03845, VR 2022-04683, VR grant 2021-03651), Knut and Alice Wallenberg Foundation (KAW 2017.0100, KAW 2018.0157, KAW 2018.0458, KAW 2019.0447); Switzerland: Swiss National Science Foundation (SNSF - PCEFP2\_194658); United Kingdom: Leverhulme Trust (Leverhulme Trust RPG-2020-004); United States of America: U.S. Department of Energy (ECA DE-AC02-76SF00515), Neubauer Family Foundation.

%
%


%
%
%
%
%
%
%

%
%
%
\printbibliography
\clearpage
\clearpage
 
\begin{flushleft}
\hypersetup{urlcolor=black}
{\Large The ATLAS Collaboration}

\bigskip

\AtlasOrcid[0000-0002-6665-4934]{G.~Aad}$^\textrm{\scriptsize 102}$,
\AtlasOrcid[0000-0002-5888-2734]{B.~Abbott}$^\textrm{\scriptsize 120}$,
\AtlasOrcid[0000-0002-1002-1652]{K.~Abeling}$^\textrm{\scriptsize 55}$,
\AtlasOrcid[0000-0001-5763-2760]{N.J.~Abicht}$^\textrm{\scriptsize 49}$,
\AtlasOrcid[0000-0002-8496-9294]{S.H.~Abidi}$^\textrm{\scriptsize 29}$,
\AtlasOrcid[0000-0002-9987-2292]{A.~Aboulhorma}$^\textrm{\scriptsize 35e}$,
\AtlasOrcid[0000-0001-5329-6640]{H.~Abramowicz}$^\textrm{\scriptsize 152}$,
\AtlasOrcid[0000-0002-1599-2896]{H.~Abreu}$^\textrm{\scriptsize 151}$,
\AtlasOrcid[0000-0003-0403-3697]{Y.~Abulaiti}$^\textrm{\scriptsize 117}$,
\AtlasOrcid[0000-0002-8588-9157]{B.S.~Acharya}$^\textrm{\scriptsize 69a,69b,m}$
% ...
\bigskip
\\

% ...
$^{29}$Physics Department, Brookhaven National Laboratory, Upton NY; United States of America.\\
% ...
$^{35}$$^{(a)}$Facult\'e des Sciences Ain Chock, Universit\'e Hassan II de Casablanca;$^{(b)}$Facult\'{e} des Sciences, Universit\'{e} Ibn-Tofail, K\'{e}nitra;$^{(c)}$Facult\'e des Sciences Semlalia, Universit\'e Cadi Ayyad, LPHEA-Marrakech;$^{(d)}$LPMR, Facult\'e des Sciences, Universit\'e Mohamed Premier, Oujda;$^{(e)}$Facult\'e des sciences, Universit\'e Mohammed V, Rabat;$^{(f)}$Institute of Applied Physics, Mohammed VI Polytechnic University, Ben Guerir; Morocco.\\
% ...
$^{49}$Fakult\"{a}t Physik , Technische Universit{\"a}t Dortmund, Dortmund; Germany.\\
% ...
$^{55}$II. Physikalisches Institut, Georg-August-Universit\"{a}t G\"ottingen, G\"ottingen; Germany.\\
% ...
$^{69}$$^{(a)}$INFN Gruppo Collegato di Udine, Sezione di Trieste, Udine;$^{(b)}$ICTP, Trieste;$^{(c)}$Dipartimento Politecnico di Ingegneria e Architettura, Universit\`a di Udine, Udine; Italy.\\
% ...
$^{102}$CPPM, Aix-Marseille Universit\'e, CNRS/IN2P3, Marseille; France.\\
% ...
$^{117}$Department of Physics, New York University, New York NY; United States of America.\\
% ...
$^{120}$Homer L. Dodge Department of Physics and Astronomy, University of Oklahoma, Norman OK; United States of America.\\
% ...
$^{151}$Department of Physics, Technion, Israel Institute of Technology, Haifa; Israel.\\
$^{152}$Raymond and Beverly Sackler School of Physics and Astronomy, Tel Aviv University, Tel Aviv; Israel.\\
% ...
$^{a}$ Also Affiliated with an institute covered by a cooperation agreement with CERN.\\
$^{b}$ Also at An-Najah National University, Nablus; Palestine.\\
$^{c}$ Also at Borough of Manhattan Community College, City University of New York, New York NY; United States of America.\\
$^{d}$ Also at Center for High Energy Physics, Peking University; China.\\
$^{e}$ Also at Center for Interdisciplinary Research and Innovation (CIRI-AUTH), Thessaloniki; Greece.\\
$^{f}$ Also at Centro Studi e Ricerche Enrico Fermi; Italy.\\
% ...
$^{m}$ Also at Department of Physics, King's College London, London; United Kingdom.\\
% ...

\end{flushleft}


%

%
%

%

%
%
%
%

\end{document}